\documentclass[12pt]{article}

\usepackage{color}
\usepackage{caption}
\usepackage{subcaption}
\usepackage{scicite}
\usepackage{url}
\usepackage{mhchem}
\usepackage{amssymb}
\usepackage{tabu}
\usepackage{hyperref}
\hypersetup{
	colorlinks=true,
	citecolor=blue
}

\usepackage{epsfig}
\usepackage[english]{babel}

\usepackage{times}

\newcounter{lastnote}

\title{Pore-geometry recognition: on the importance of quantifying similarity in nanoporous materials}

\author
{Yongjin~Lee,$^{1,2}$ Senja~D.~Barthel,$^{1}$ Pawe\l{}~D\l{}otko,$^{3}$ S.~Mohamad~Moosavi,$^{1}$\\Kathryn Hess,$^{4}$ and Berend Smit$^{1,2\ast}$ \\
\\
\normalsize{$^{1}$Institut des Sciences et Ing\'{e}nierie Chimiques, Valais, Ecole Polytechnique F\'{e}d\'{e}rale de Lausanne (EPFL),}\\
\normalsize{Rue de l'Industrie 17, CH-1951 Sion, Switzerland }\\
\normalsize{$^{2}$Department of Chemical and Biomolecular Engineering, University of California,}\\
\normalsize{Berkeley, CA 94720, USA}\\
\normalsize{$^{3}$DataShape Group, Inria Saclay – Ile-de-France, 91120 Palaiseau, France}\\
\normalsize{$^{3}$SV BMI UPHESS, Ecole Polytechnique F\'{e}d\'{e}rale de Lausanne, CH-1015 Lausanne, Switzerland}\\
\\
\normalsize{$^\ast$berend.smit@epfl.ch}
}

\date{}

\begin{document} 


\baselineskip24pt

\maketitle 



{\bf
In most applications of nanoporous materials the pore structure is as important as the chemical composition as a determinant of performance. For example, one can alter performance in applications like carbon capture or methane storage by orders of magnitude by only modifying the pore structure.\cite{sim141,lin121} For these applications it is therefore important to identify the optimal pore geometry and use this information to find similar materials. However, the mathematical language and tools to identify materials with similar pore structures, but different composition, has been lacking. Here we develop a pore recognition approach to quantify similarity of pore structures and classify them using topological data analysis.\cite{ede101,car091} Our approach allows us to identify materials with similar pore geometries, and to screen for materials that are similar to given top-performing structures. Using methane storage as a case study, we also show that materials can be divided into topologically distinct classes -- and that each class requires different optimization strategies. In this work we have focused on pore space, but our topological approach can be generalised to quantify similarity of any geometric object, which, given the many different Materials Genomics initiatives,\cite{nos161,kal111} opens many interesting avenues for big-data science.
}

Understanding Big Data is a challenge social and natural sciences share. The need to handle huge amounts of data, often generated by the steady increase of available computing power, has inspired rapid development in big-data science. In chemistry and material science, new research initiatives (e.g., the materials genome initiative\cite{kal111}) have led to the generation of large databases of materials for different applications. 

We focus on nanoporous materials, such as zeolites,\cite{pop111} metal organic frameworks (MOFs),\cite{fur131} zeolitic imidizolate frameworks (ZIFs),\cite{par061} and porous polymer networks (PPNs).\cite{coo091} These materials are of interest in applications ranging from gas separation and storage, to catalysis. In each case one would like to tailor-make a material that is optimal for that particular application. The chemistry of these materials allows us to obtain an essentially unlimited number of new materials.\cite{wil121,chu141,bao151,mar131,mar141} Indeed,  in recent years the number of published synthesized nanoporous materials has grown exponentially.\cite{fur131} Yet, this growth is exceeded by the number of predicted structures, giving us libraries of millions of potentially interesting new materials. This sheer abundance of structures requires novel techniques from big-data research to shed light on the existing libraries, as well as to facilitate the search for materials with optimal properties.

In nanoporous materials the shape of the pores plays an essential role in the behavior of the material. Conventionally, pore structure is characterized by a set of traditional geometric descriptors such as pore volume, largest included sphere, surface area, etc. These descriptors can be successfully optimized to search for materials with similar overall thermodynamic properties, but, as we will show, they capture partial geometric features only and do not encode enough geometric information to enable detection of materials that have similar overall pore shapes. There are computational techniques to quantify the similarity between crystal structures.\cite{zhu161,oga091} However, these algorithms are limited to identifying identical crystal structures, while we are interested in finding materials that may have different crystal structures or chemical compositions but similar pore geometries. Martin {\it et al}.\cite{mar121} developed Voronoi network representations of pore geometries, which are useful as fingerprints but do not capture details of the local pore structure. In this Letter, we develop a mathematical quantification of geometric similarity by using topological data analysis (TDA). TDA is a field of big-data analysis that builds on techniques from algebraic topology, most noticeably persistent homology.\cite{ede101} Its guiding philosophy is that the `shape' of the data reveals important information about the data.\cite{car091} 

To assign a geometric descriptor to a given material, we sample points on the pore surface. By growing balls stepwise around each sample point and monitoring their pairwise overlaps, we compute the associated filtered Vietoris-Rips complex (see SI section~\ref{theo}), which is then characterized by its 0-, 1- and 2-dimensional homology classes. We store the lifetime of each homology class in the corresponding persistence barcode. Combining the 0-, 1-, and 2-dimensional barcodes yields a fingerprint that characterizes the overall shape of the pore structure. 

\begin{figure}[h!]
\begin{center}
\includegraphics[width=\textwidth]{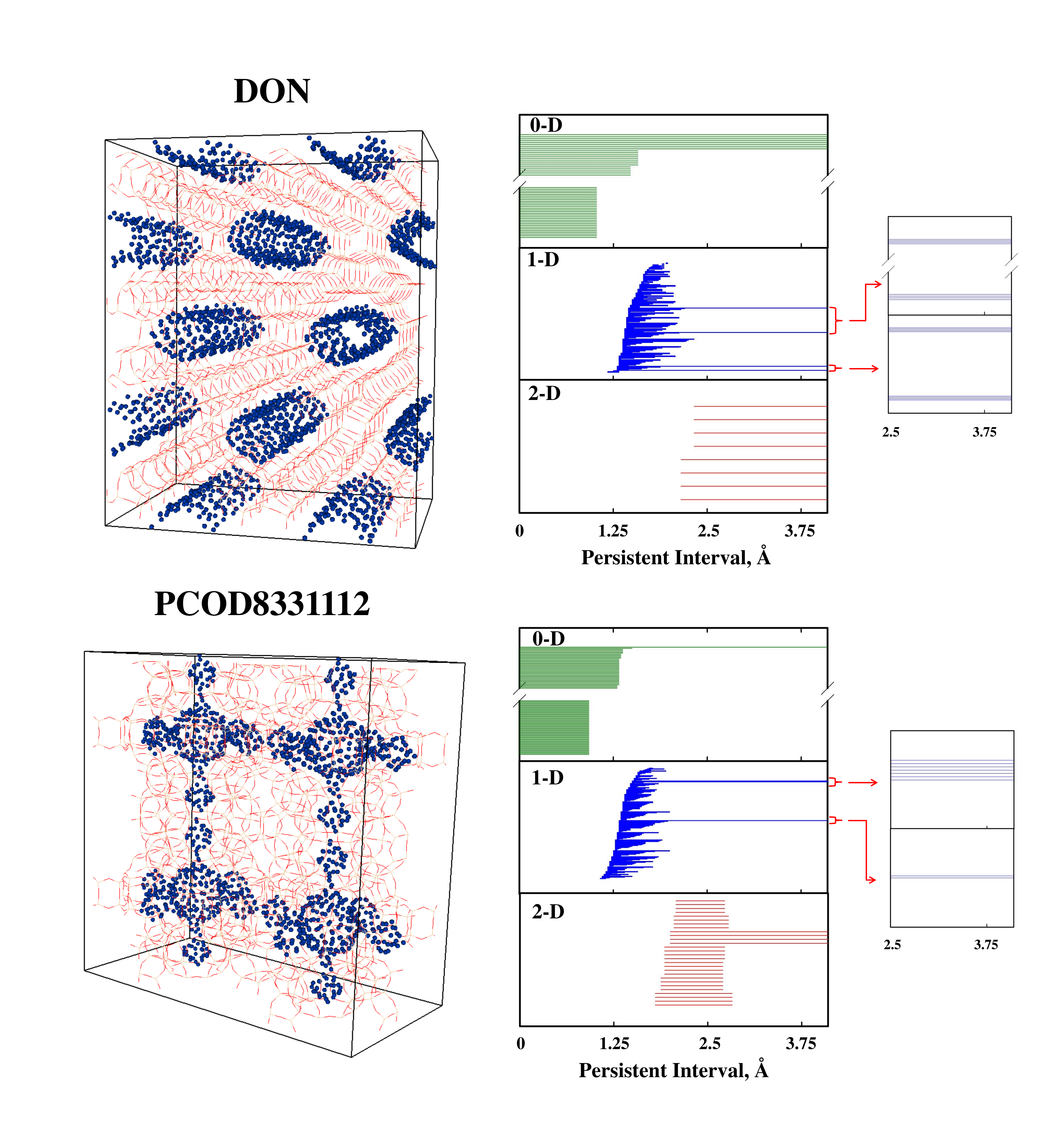}
\end{center}
\vskip -1cm
\caption{}
\label{DONh}
\end{figure}

For analyzing pore shapes we are in the unusually fortunate situation that, unlike most other big-data applications of persistent homology, our data have actual geometric meaning. In almost all known big-data applications only the 0-D and 1-D barcodes are of relevance, while here the 2-D barcodes also carry essential information. For example, figure~\ref{DONh} shows the fingerprints of two different zeolite structures, IZA zeolite DON and hypothetical zeolite PCOD8331112. DON contains eight identical cylindrical pores that run parallel to each other. The pore structure of PCOD8331112 is a 3-dimensional network that is formed of two types of connected spherical cavities. The 0-D barcodes of both structures start with as many intervals as there are points sampled on the pore surfaces. More information is contained in the long intervals describing robust shape features: the existence of the single long interval in its 0-D barcode implies that the pore system of PCOD8331112 is connected. In contrast, the pore system of DON consists of eight unconnected components, encoded by the eight long intervals in its 0-D barcode. The 1-D and 2-D barcodes contain information on the shape of the cavities (see SI section~\ref{theo} for details). 

\begin{figure}
\centering

\includegraphics[width=0.9\textwidth]{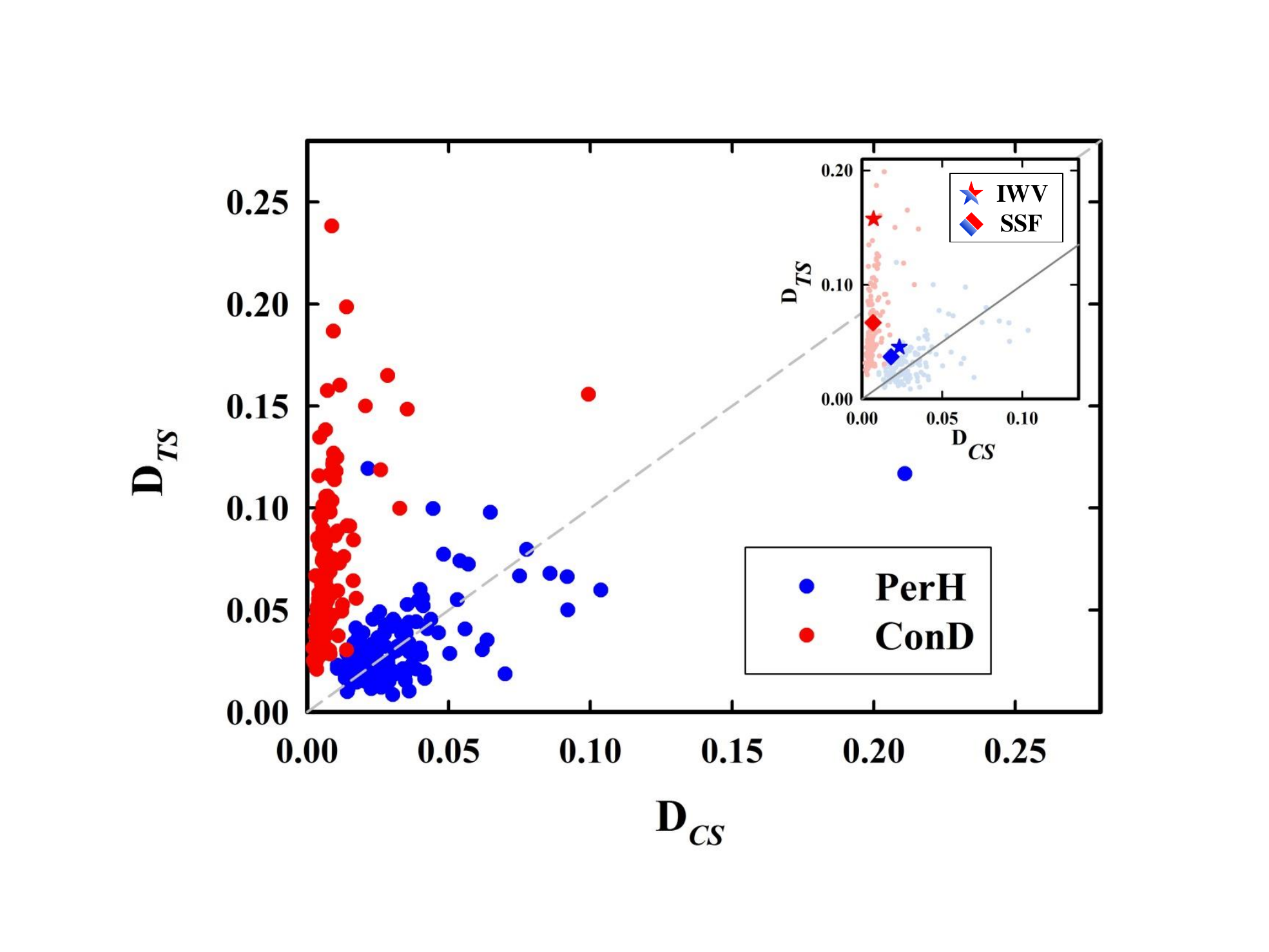}\\
(a) \\

\includegraphics[width=\textwidth]{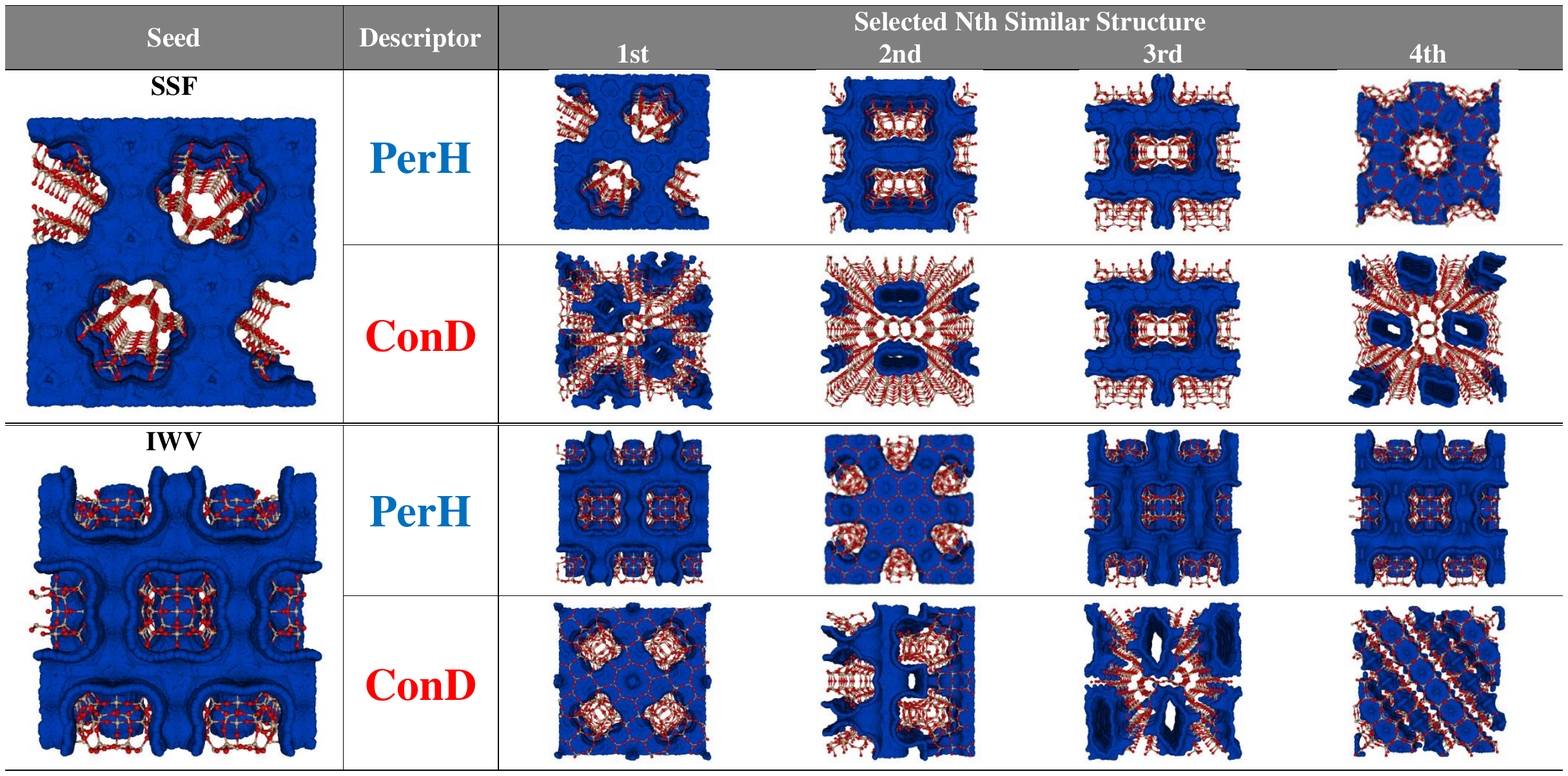}
\vskip -2.5cm
(b)\\
\caption{}
\label{Compare}\label{4similar}
\end{figure}

The most elementary, but highly non-trivial, application of our approach is to identify porous materials with similar pore structures. As we have a database of over 3,000,000 nanoporous structures,\cite{sim151} visual inspection is out of question. Suppose we would like to know whether the library of hypothetical zeolites contains structures whose pore geometry is similar to a given material, e.g., a synthesized zeolite. To see the effectiveness of our approach, it is instructive to take a structure and find the four structures that are most similar to the chosen one, selected once by conventional descriptors (ConD) and once using persistent homology (PerH). To compare these two sets, we compute their average distances to the reference material, measured by the metric $D_{CS}$ of the conventional space as well as by the metric $D_{TS}$ of the barcode space (see SI~sections~\ref{methcomp} and \ref{theo} for details). Figure~\ref{Compare}a shows the average distances of the two sets for each of the 146 experimentally known zeolite structures accessible to methane taken. The distances are normalized by the largest pairwise distance in the database. The TDA approach provides what one would expect: when persistent homology is used to identify similar pore structures, small $D_{TS}$ correlates well with small $D_{CS}$. I.e., similar persistence diagrams describing the pore shapes correlate with similar conventional geometric measures. Figure~\ref{4similar}a shows that the relatively few zeolites for which there are no four structures very similar to a given one with respect to PerH (large $D_{TS}$), the first four structures chosen by PerH might or might not have similar conventional geometric descriptors (small or large $D_{CS}$).
The conventional approach, however, gives a different result: for each reference structure we can find structures with similar conventional descriptors (small $D_{CS}$) but the shapes of their pores can differ enormously (large $D_{TS}$). Figure~\ref{4similar}b shows two cases where the conventional approach identifies structures with very similar conventional descriptors (see SI Table~\ref{StructProp}) but very different pore structures. In contrast, if we use our topology-based fingerprint, we indeed retrieve structures that look strikingly similar. In the SI (section~\ref{similarity}) we show that one can also use this similarity search to compare structures from different classes of nanoporous materials. These findings are guaranteed by a stability theorem that is a key result in persistent homology:\cite{cha141} materials with similar shapes are described by similar barcodes.

For the traditional descriptors with geometric meaning, one expects to find correlations with information encoded in the persistent homological fingerprint (see SI Table~\ref{BarcodeErrors}). For example, the radius of the maximum included sphere is correlated with the 2-D barcode as the radius of a cavity determines the death time of an interval in the 2-D-barcode. Further geometric information, like the connectivity of the pore structure (0-D) or the number of independent tunnels (1-D), is also encoded in the persistence barcodes. Therefore, only the combination of the barcodes of all three dimensions captures the global geometric features of the pore shapes we are interested in.

One of the characteristics of MOFs is their chemical tunability. Indeed, over the last 5 years over 10,000 structures have been synthesised.\cite{fur131} Such a large number of materials makes it simply impossible to know all corresponding pore structures by heart. Therefore, an important application of our methodology is that we can now readily identify similar pore structures. In Figure~\ref{f:coreMOFs} we show some examples of materials from the CoRE-MOF database  that have similar pore geometries. Our list of similar structures in much longer but what is specific to these examples is that the authors of the corresponding manuscripts did not report the similarities in the original references. Of course, this does not imply that the authors of these articles were not aware of these similarities, but given that there are over 10,000 experimental MOF structures,  such similarities are easily overlooked.

An important practical application of nanoporous materials is methane storage. The performance property of this application is deliverable capacity, which is defined as the difference between the amount of methane that is adsorbed at the (high) pressure at which the material is charged and the amount that remains in the material at the de-charging (low) pressure; the higher this deliverable capacity, the better the material. One of the interesting features of nanoporous materials is that one can optimize the pore geometry for a given application. The idea is that if one identifies a material with a high deliverable capacity, materials with similar pore geometries should also have an excellent performance. Let us illustrate this idea using all-silica zeolites. For this class of nanoporous materials the chemical composition (Si/O) is the same, hence the determining factor is the pore shape. From molecular simulations we have determined the 13 best performing out of the 180 know-zeolite structures, each having a deliverable capacity larger that 90 (v STP/v). We subsequently identified for each of these top-performing materials the 10 most similar structures in our database of 139,407 predicted zeolites. Figure~\ref{f:CH4}~(a) shows that indeed 80\% of these 130 new structures have a deliverable capacity that is similar to the 13 top-performing knowns-zeolites. In Figure~\ref{f:CH4}~(b) we show a similar results for MOFs where we used the 20 top-performing structures from the CoRE-MOF database and identified similar structure in the databases of 41498 predicted MOF structures: 85\% of these materials show high performance with a deliverable capacity larger than 150 (v STP/v). It is interesting that even for MOFs which have different chemical compositions (unlike zeolites), our method of identifying similar pore shapes illustrates the importance of pore geometry, and hence, of our methodology to quantify similarity for these types of materials.

We can also use our approach to study the topological diversity of the top-performing materials for methane storage. Bathia and Myers\cite{bha061} analyzed a small number of porous materials and concluded that top-performing materials should all have very similar heats of adsorption for given loading and de-charging pressures. Their work has had significant impact, as it provides a straightforward experimental recipe for optimizing the deliverable capacity of a material.\cite{mas141} I.e., if all top-performing materials shared a similar heat of adsorption, having a heat of adsorption similar to this value was a necessary condition for all good performing materials. Given this impact it is surprising that the conclusion of Bathia and Myers stands in sharp contrast with observations of Simon {\it et al}.\cite{sim141} Simon {\it et al.} computed the deliverable capacity for over 200,000 zeolite structures, and their data (Figure~\ref{QadGroups}a) provide no evidence for a single optimal heat of adsorption, pointing to an interesting paradox: if one randomly selects a set of materials from Figure~\ref{QadGroups}a, one finds no experimental indication that an optimal heat of adsorption even exists. Yet, the approach of Bathia and Myers has indeed been shown to be useful in optimizing performance.

To shed some light on this paradox, we applied topological data analysis to the data in Figure~\ref{QadGroups}a.  Analyzing the heat of adsorption for sets of geometrically similar structures, we obtain the desired `volcano plots' shown in Figure~\ref{QadGroups}b, which allow us to systematically search for the optimal heat of adsorption within a class of geometrically similar structures, and hence the best-performing materials. Interestingly, this optimal heat of adsorption depends on the geometric type of a material\cite{bae101,fro071} (see Figure~\ref{QadGroups}) and is not, as suggested by Bathia and Myers, a universal constant. In fact, Bathia and Myers assume implicitly that the entropy of adsorption is the same for all materials; for a set of similar materials as often chosen this assumption is more likely to hold. 
 
The results above suggest that there is not a single class of optimal materials. For this particular example Simon {\it et al}.\cite{sim141} have used brute-force simulations to compute the performance of all materials. This allows us to use the TDA  approach to analyze the geometric diversity of the top-performing structures, and to visualize the topography of the zeolite library by generating the mapper plot\cite{car091,lum131} shown in Figure~\ref{mapper}, encoding the topological structure of the set of the best 1\% performing zeolites with respect to methane storage. 
  
The shape of the diagram shows seven topologically different classes of top-performing materials. For example, group C consists of materials that have one-dimensional small cylinders, while group E has two-dimensional channels (see Table~\ref{Groups} for all different groups).  
The color coding of the mapper plot nicely illustrates that materials in classes of different pore shapes have very different optimal heats of adsorption. 

In this article we have developed a topology-based methodology to quantify similarity of the chemical environment of adsorbed molecules, focusing on applications in which the pores play a passive role in providing adsorption sites. For applications in which the pores play a more active role, such as catalysis, a logical step would be to extend the methodology to include chemical specificity and charge distribution. From a topology viewpoint this application  is of  particular significance because it is one of the first applications of topological data analysis that requires persistent homology in three different dimensions.

\baselineskip14pt
\bibliography{nature} 

\begin{thebibliography}{10}
\expandafter\ifx\csname url\endcsname\relax
  \def\url#1{\texttt{#1}}\fi
\expandafter\ifx\csname urlprefix\endcsname\relax\def\urlprefix{URL }\fi
\providecommand{\bibinfo}[2]{#2}
\providecommand{\eprint}[2][]{\url{#2}}

\bibitem{sim141}
\bibinfo{author}{SImon, C.~M.} \emph{et~al.}
\newblock \bibinfo{title}{Optimizing nanoporous materials for gas storage}.
\newblock \emph{\bibinfo{journal}{Phys. Chem. Chem. Phys.}}
  \textbf{\bibinfo{volume}{16}}, \bibinfo{pages}{5499--5513}
  (\bibinfo{year}{2014}).

\bibitem{lin121}
\bibinfo{author}{Lin, L.-C.} \emph{et~al.}
\newblock \bibinfo{title}{In silico screening of carbon-capture materials}.
\newblock \emph{\bibinfo{journal}{Nat. Mater.}} \textbf{\bibinfo{volume}{11}},
  \bibinfo{pages}{633--641} (\bibinfo{year}{2012}).

\bibitem{ede101}
\bibinfo{author}{Edelsbrunner, H.} \& \bibinfo{author}{Harer, J.}
\newblock \emph{\bibinfo{title}{Computational topology: an introduction}}
  (\bibinfo{publisher}{American Mathematical Soc.}, \bibinfo{year}{2010}).

\bibitem{car091}
\bibinfo{author}{Carlsson, G.}
\newblock \bibinfo{title}{Topology and data}.
\newblock \emph{\bibinfo{journal}{Bull. Amer. Math. Soc.}}
  \textbf{\bibinfo{volume}{46}}, \bibinfo{pages}{255--308}
  (\bibinfo{year}{2009}).

\bibitem{nos161}
\bibinfo{author}{Nosengo, N.}
\newblock \bibinfo{title}{The material code}.
\newblock \emph{\bibinfo{journal}{Nature}} \textbf{\bibinfo{volume}{533}},
  \bibinfo{pages}{22--25} (\bibinfo{year}{2016}).

\bibitem{kal111}
\bibinfo{author}{Kalil, T.} \& \bibinfo{author}{Wadia, C.}
\newblock \bibinfo{title}{Materials genome initiative for global
  competitiveness} (\bibinfo{year}{2011}).

\bibitem{pop111}
\bibinfo{author}{Pophale, R.}, \bibinfo{author}{Cheeseman, P.~A.} \&
  \bibinfo{author}{Deem, M.~W.}
\newblock \bibinfo{title}{A database of new zeolite-like materials}.
\newblock \emph{\bibinfo{journal}{Phys Chem Chem Phys}}
  \textbf{\bibinfo{volume}{13}}, \bibinfo{pages}{12407--12412}
  (\bibinfo{year}{2011}).

\bibitem{fur131}
\bibinfo{author}{Furukawa, H.}, \bibinfo{author}{Cordova, K.~E.},
  \bibinfo{author}{O'Keeffe, M.} \& \bibinfo{author}{Yaghi, O.~M.}
\newblock \bibinfo{title}{The chemistry and applications of metal-organic
  frameworks}.
\newblock \emph{\bibinfo{journal}{Science}} \textbf{\bibinfo{volume}{341}},
  \bibinfo{pages}{974} (\bibinfo{year}{2013}).

\bibitem{par061}
\bibinfo{author}{Park, K.~S.} \emph{et~al.}
\newblock \bibinfo{title}{Exceptional chemical and thermal stability of
  zeolitic imidazolate frameworks}.
\newblock \emph{\bibinfo{journal}{Proc. Nat. Acad. Sci. U.S.A.}}
  \textbf{\bibinfo{volume}{103}}, \bibinfo{pages}{10186--10191}
  (\bibinfo{year}{2006}).

\bibitem{coo091}
\bibinfo{author}{Cooper, A.~I.}
\newblock \bibinfo{title}{Conjugated microporous polymers}.
\newblock \emph{\bibinfo{journal}{Adv Mater}} \textbf{\bibinfo{volume}{21}},
  \bibinfo{pages}{1291--1295} (\bibinfo{year}{2009}).

\bibitem{wil121}
\bibinfo{author}{Wilmer, C.~E.} \emph{et~al.}
\newblock \bibinfo{title}{Large-scale screening of hypothetical metal organic
  frameworks}.
\newblock \emph{\bibinfo{journal}{Nat Chem}} \textbf{\bibinfo{volume}{4}},
  \bibinfo{pages}{83--89} (\bibinfo{year}{2012}).

\bibitem{chu141}
\bibinfo{author}{Chung, Y.~G.} \emph{et~al.}
\newblock \bibinfo{title}{Computation-ready, experimental metal-organic
  frameworks: A tool to enable high-throughput screening of nanoporous
  crystals}.
\newblock \emph{\bibinfo{journal}{Chem Mater}} \textbf{\bibinfo{volume}{26}},
  \bibinfo{pages}{6185--6192} (\bibinfo{year}{2014}).

\bibitem{bao151}
\bibinfo{author}{Bao, Y.} \emph{et~al.}
\newblock \bibinfo{title}{In silico discovery of high deliverable capacity
  metal organic frameworks}.
\newblock \emph{\bibinfo{journal}{J. Phys. Chem. C}}
  \textbf{\bibinfo{volume}{119}}, \bibinfo{pages}{186--195}
  (\bibinfo{year}{2015}).

\bibitem{mar131}
\bibinfo{author}{Martin, R.~L.}, \bibinfo{author}{Lin, L.-C.},
  \bibinfo{author}{Jariwala, K.}, \bibinfo{author}{Smit, B.} \&
  \bibinfo{author}{Haranczyk, M.}
\newblock \bibinfo{title}{Mail-order metal–organic frameworks (mofs):
  Designing isoreticular mof-5 analogues comprising commercially available
  organic molecules}.
\newblock \emph{\bibinfo{journal}{J Phys Chem C}}
  \textbf{\bibinfo{volume}{117}}, \bibinfo{pages}{12159−12167}
  (\bibinfo{year}{2013}).

\bibitem{mar141}
\bibinfo{author}{Martin, R.~L.}, \bibinfo{author}{Simon, C.~M.},
  \bibinfo{author}{Smit, B.} \& \bibinfo{author}{Haranczyk, M.}
\newblock \bibinfo{title}{In silico design of porous polymer networks:
  High-throughput screening for methane storage materials}.
\newblock \emph{\bibinfo{journal}{J. Am. Chem. Soc.}}
  \textbf{\bibinfo{volume}{136}}, \bibinfo{pages}{5006--5022}
  (\bibinfo{year}{2014}).

\bibitem{zhu161}
\bibinfo{author}{Zhu, L.} \emph{et~al.}
\newblock \bibinfo{title}{A fingerprint based metric for measuring similarities
  of crystalline structures}.
\newblock \emph{\bibinfo{journal}{J Chem Phys}} \textbf{\bibinfo{volume}{144}}
  (\bibinfo{year}{2016}).

\bibitem{oga091}
\bibinfo{author}{Oganov, A.~R.} \& \bibinfo{author}{Valle, M.}
\newblock \bibinfo{title}{How to quantify energy landscapes of solids}.
\newblock \emph{\bibinfo{journal}{J Chem Phys}} \textbf{\bibinfo{volume}{130}}
  (\bibinfo{year}{2009}).

\bibitem{mar121}
\bibinfo{author}{Martin, R.~L.}, \bibinfo{author}{Smit, B.} \&
  \bibinfo{author}{Haranczyk, M.}
\newblock \bibinfo{title}{Addressing challenges of identifying geometrically
  diverse sets of crystalline porous materials}.
\newblock \emph{\bibinfo{journal}{J. Chem Inf. Model.}}
  \textbf{\bibinfo{volume}{52}}, \bibinfo{pages}{308--318}
  (\bibinfo{year}{2012}).

\bibitem{sim151}
\bibinfo{author}{Simon, C.} \emph{et~al.}
\newblock \bibinfo{title}{The materials genome in action: Identifying the
  performance limits for methane storage}.
\newblock \emph{\bibinfo{journal}{Energy Environ. Sci.}}
  \textbf{\bibinfo{volume}{8}}, \bibinfo{pages}{1190–1199}
  (\bibinfo{year}{2015}).

\bibitem{cha141}
\bibinfo{author}{Chazal, F.}, \bibinfo{author}{de~Silva, V.} \&
  \bibinfo{author}{Oudot, S.}
\newblock \bibinfo{title}{Persistence stability for geometric complexes}.
\newblock \emph{\bibinfo{journal}{Geometriae Dedicata}}
  \textbf{\bibinfo{volume}{173}}, \bibinfo{pages}{193--214}
  (\bibinfo{year}{2014}).

\bibitem{bha061}
\bibinfo{author}{Bhatia, S.~K.} \& \bibinfo{author}{Myers, A.~L.}
\newblock \bibinfo{title}{Optimum conditions for adsorptive storage}.
\newblock \emph{\bibinfo{journal}{Langmuir}} \textbf{\bibinfo{volume}{22}},
  \bibinfo{pages}{1688--1700} (\bibinfo{year}{2006}).

\bibitem{mas141}
\bibinfo{author}{Mason, J.~A.}, \bibinfo{author}{Veenstra, M.} \&
  \bibinfo{author}{Long, J.~R.}
\newblock \bibinfo{title}{Evaluating metal-organic frameworks for natural gas
  storage}.
\newblock \emph{\bibinfo{journal}{Chem. Scie.}} \textbf{\bibinfo{volume}{5}},
  \bibinfo{pages}{32--51} (\bibinfo{year}{2014}).

\bibitem{bae101}
\bibinfo{author}{Bae, Y.~S.} \& \bibinfo{author}{Snurr, R.~Q.}
\newblock \bibinfo{title}{Optimal isosteric heat of adsorption for hydrogen
  storage and delivery using metal-organic frameworks}.
\newblock \emph{\bibinfo{journal}{Micropor Mesopor Mat}}
  \textbf{\bibinfo{volume}{132}}, \bibinfo{pages}{300--303}
  (\bibinfo{year}{2010}).

\bibitem{fro071}
\bibinfo{author}{Frost, H.} \& \bibinfo{author}{Snurr, R.~Q.}
\newblock \bibinfo{title}{Design requirements for metal-organic frameworks as
  hydrogen storage materials}.
\newblock \emph{\bibinfo{journal}{J Phys Chem C}}
  \textbf{\bibinfo{volume}{111}}, \bibinfo{pages}{18794--18803}
  (\bibinfo{year}{2007}).

\bibitem{lum131}
\bibinfo{author}{Lum, P.~Y.} \emph{et~al.}
\newblock \bibinfo{title}{Extracting insights from the shape of complex data
  using topology}.
\newblock \emph{\bibinfo{journal}{Sci Rep}} \textbf{\bibinfo{volume}{3}}
  (\bibinfo{year}{2013}).

\end{thebibliography}
\bibliographystyle{naturemag}

\clearpage
\baselineskip24pt


{\bf Acknowledgements} This project has received funding from the European Research Council (ERC) under the European Union's Horizon 2020 research and innovation program (grant agreement No 666983 and No 339025), the National Center of Competence in Research (NCCR) ` Materials' Revolution: Computational Design and Discovery of Novel Materials (MARVEL)' of the Swiss National Science Foundation (SNSF), and the Center for Gas Separations Relevant to Clean Energy Technologies, an Energy Frontier Research Center funded by the DOE, Office of Science, Office of Basic Energy Sciences under award DE-SC0001015 (a detailed acknowledgment is given in the SI).

{\bf Additional information} Supplementary information is available in the online version of the paper. Reprints and permissions information is available at www.nature.com
/reprints.  Readers are welcome to comment on the online version of the paper. Correspondence and requests for materials should be addressed to B.S.(berend.smit@epfl.ch).

{\bf Competing financial interests} The authors declare no competing financial interests.


\clearpage

\clearpage
\section*{Figure legends}

Fig.\ 1. {\bf Examples of fingerprints obtained by persistent homology for two different zeolite structures DON (top) and PCOD8331112 (bottom).} The figures on the left show the structures, the middle the fingerprints, and the right magnifies details of the 1-D fingerprints. The red lines in the figures on the left show the zeolite structures and the navy dots are the set of randomly sampled points on the pore surfaces. The SI contains animations of growing these fingerprints.
\vskip 2pt  
~\\
Fig.\ 2. {\bf Structures similar to a reference material}  (a) For each known zeolite, the two sets of four most similar structures, once  selected using the TDA descriptor (PerH, one blue dot for a set of four) and once selected by the conventional descriptor (ConD, one red dot per set) are compared. This is done by plotting their average distances $D_{CS}$ in conventional space (x-axis) and their average distance $D_{TS}$ in the barcode space (y-axis) to the reference zeolite. The distances are normalized by the largest pairwise distance in the database. (b) The four structures most similar to the zeolite SSF respectively to IWV, as selected by either PerH or ConD. Their structural properties are given in Table~\ref{StructProp}. The inset in (a) highlights the four sets of four structures shown in (b).
\vskip 2pt  
~\\
~\\Fig.\ 3. {\bf Materials from the CoRE-MOF database that have a similar pore geometry}. Each row gives examples of materials that are very similar. There are many more similar structures in the CoRE-MOF data base than we have listed here. The ones that are listed are those in which there are no cross references in the original articles of the corresponding similar structures. 
\vskip 2pt  

~\\
~\\Fig.\ 4. {\bf Deliverable capacity of the 10 materials that are most similar to the best performing 13~zeolites (a) respectively 20~MOFs (b) with respect to PerH}.
\vskip 2pt  

~\\Fig.\ 5. {\bf The deliverable capacity and heat of adsorption of zeolites}. (a) The deliverable capacity and heat of adsorption of all zeolites (data from Simon {\it et al.}~({\it 1})). 
(b) Four reference structures IFR, LEV, VSV  and BIK were chosen and for each of them we show the 500 geometrically most similar materials (with respect to our topological descriptor) highlighted on the plot from (a). The optimal heats of adsorptions for these subsets are depicted with the vertical lines in (a).
\vskip 2pt  
~\\
Fig.\ 6. {\bf Mapper plot generated by performing Topological Data Analysis (TDA) on the subgroups of top-performing zeolites (top 1\%) for methane storage application.} Nodes in the network represent clusters of materials with similar pore shapes and edges connect nodes that contain structures in common. Each material is represented by its persistent barcodes and the metric in this space is $D_{TS}$. Examples of the different groups are shown in SI. The inset distinguishes the top-performing materials in the training set (white circles) among all top-performing materials. These figures were obtained with the Ayasdi Core software platform (ayasdi.com). Lens: Neighborhood lens (Resolution 30, Gain 3.0x), see Ayasdi manual. Nodes are colored by the average value of the heats of adsorption of the materials in a cluster (Red: high value, Blue: low value).

\clearpage
\section*{FIGURES}

\newpage

\begin{figure}
\centering
\includegraphics[scale=0.8]{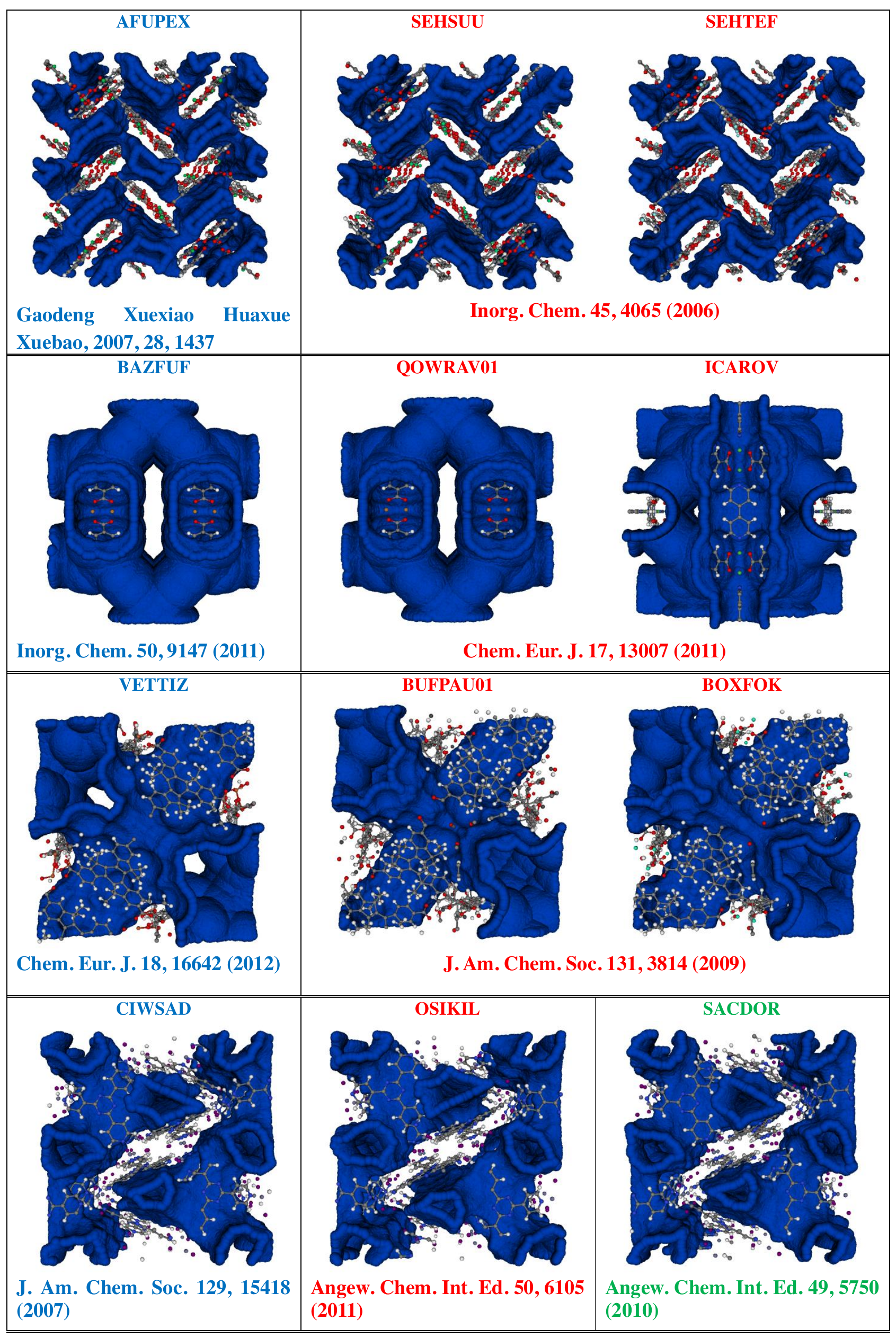}
\end{figure}
\begin{figure}
\centering
\vskip -7cm
\includegraphics[scale=0.8]{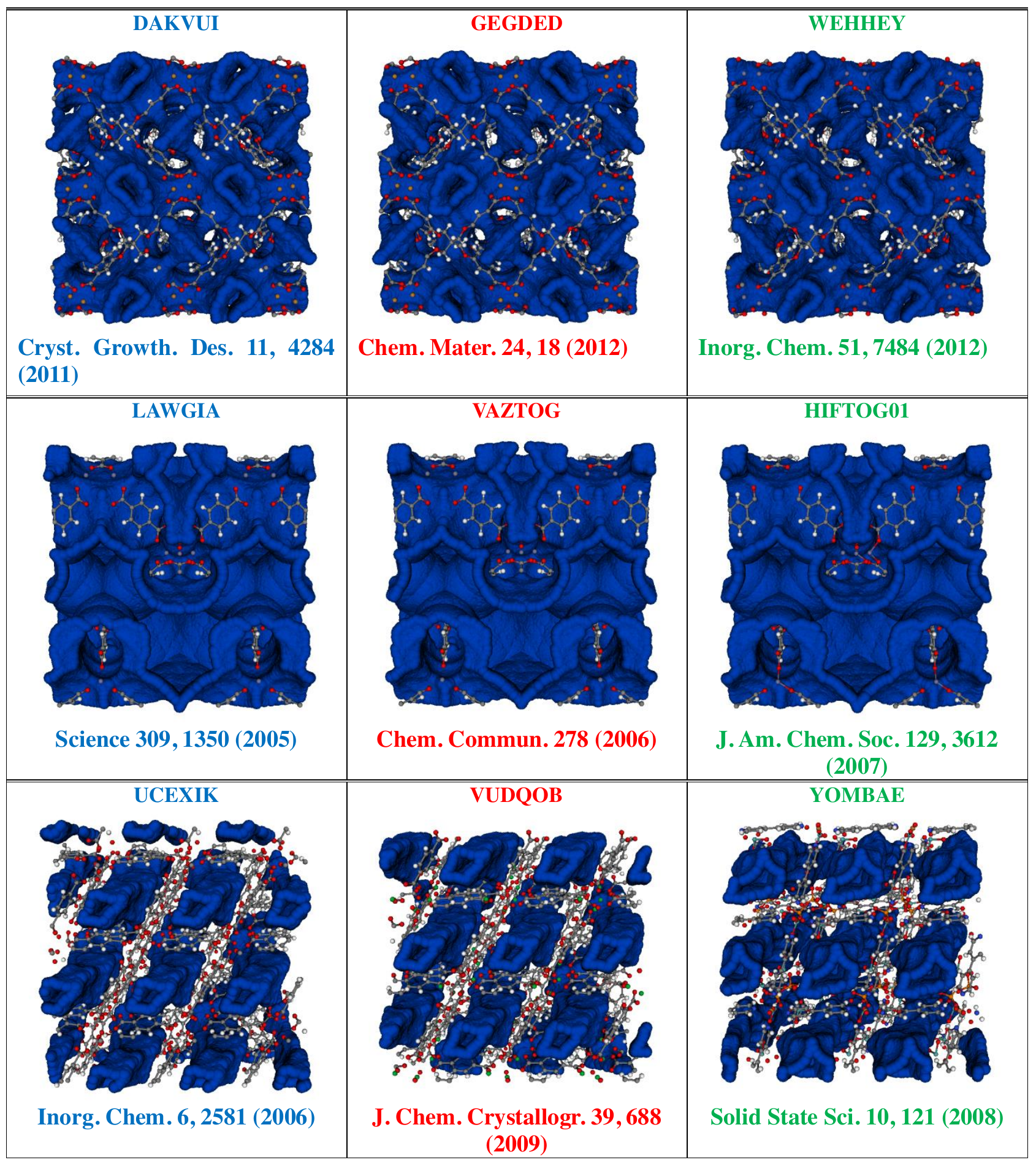}
\caption{\label{f:coreMOFs}  } 
\end{figure}
\clearpage
\begin{figure}[h!]
\begin{center}
(a)\\~\\
\includegraphics[width=12cm]{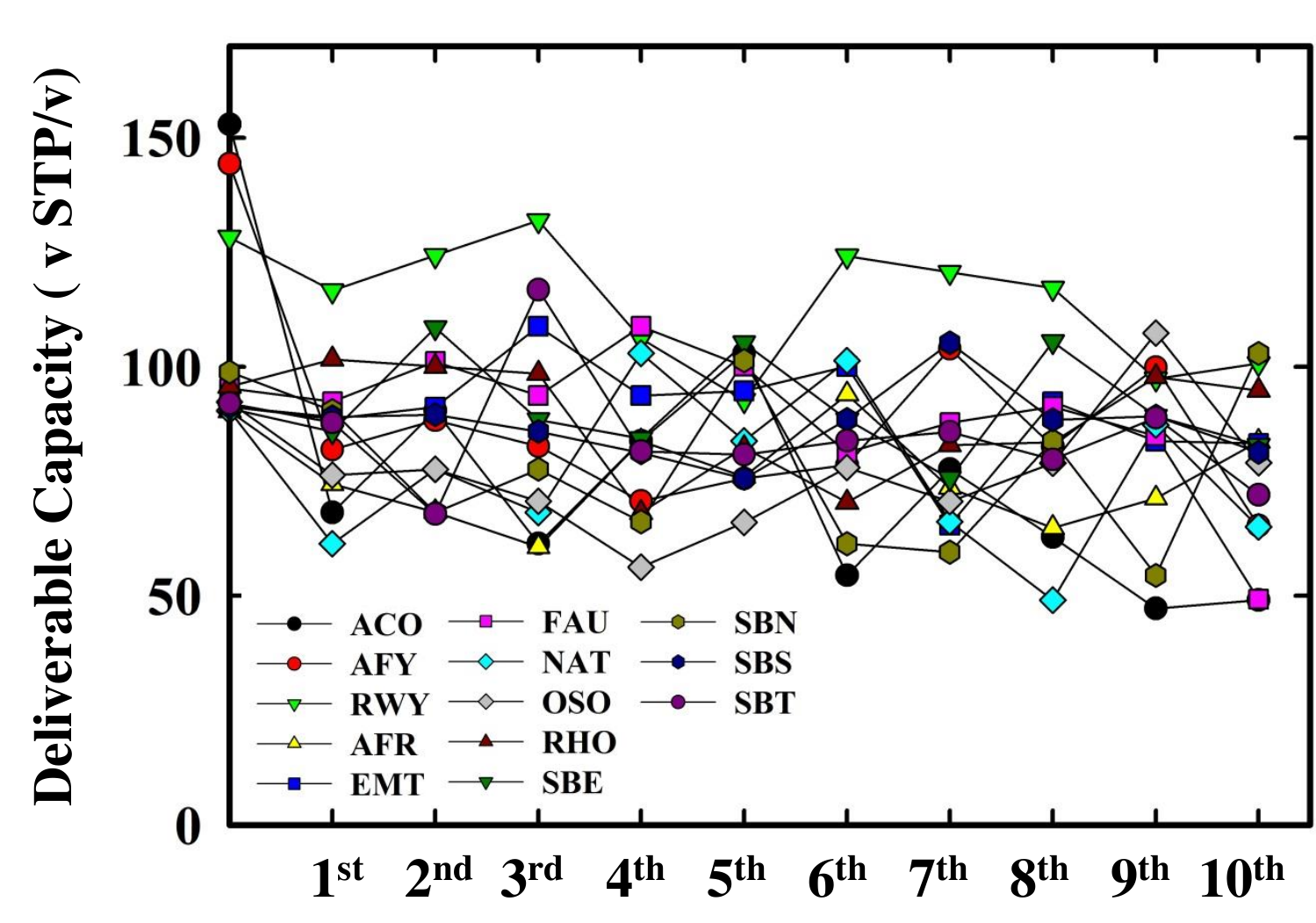}\\
~\\(b)\\~\\
\includegraphics[width=12cm]{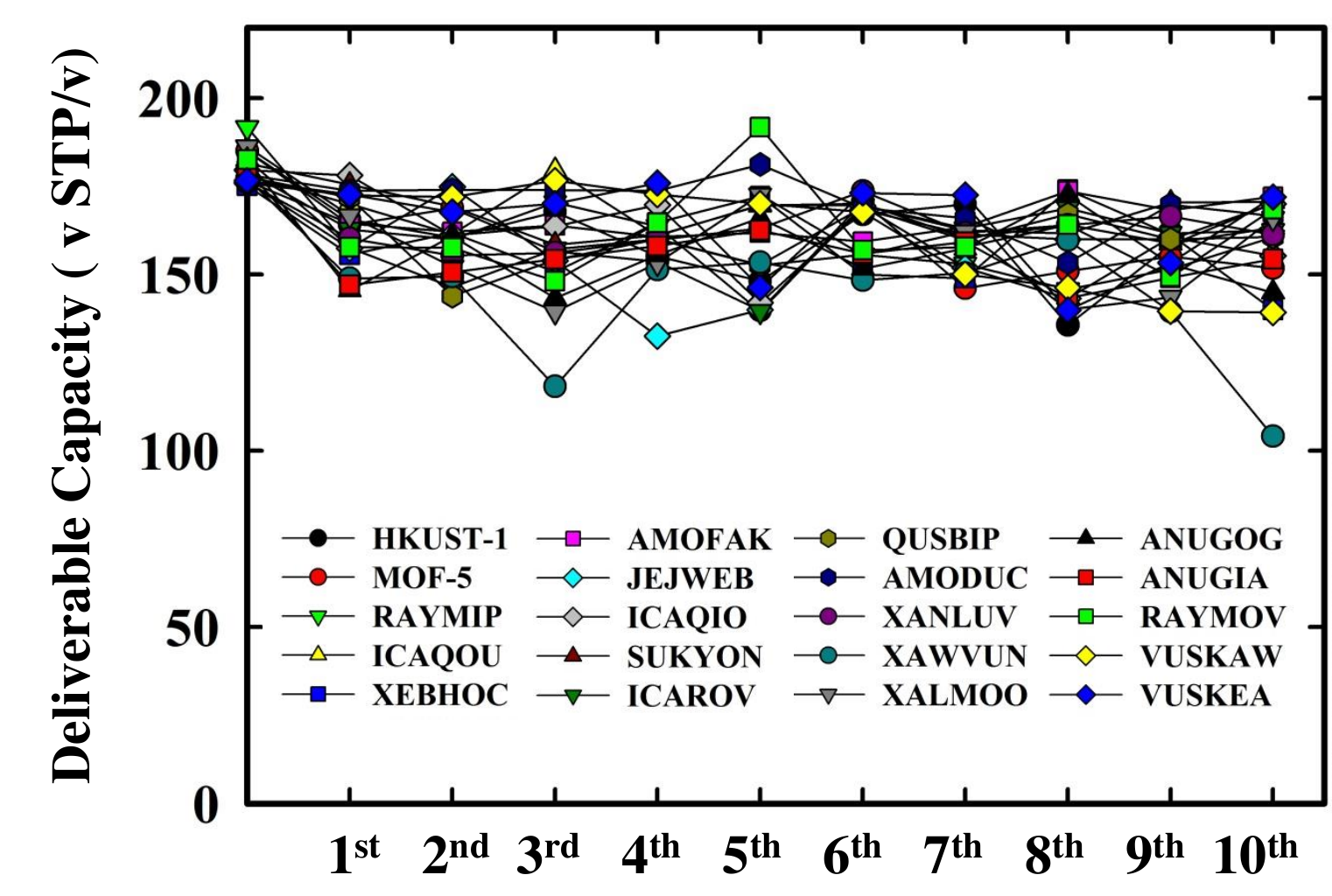}
\end{center}
\vskip 1cm
\caption{}
\label{f:CH4}
\end{figure}

\begin{figure}[h!]
\centering
\includegraphics[scale=0.5]{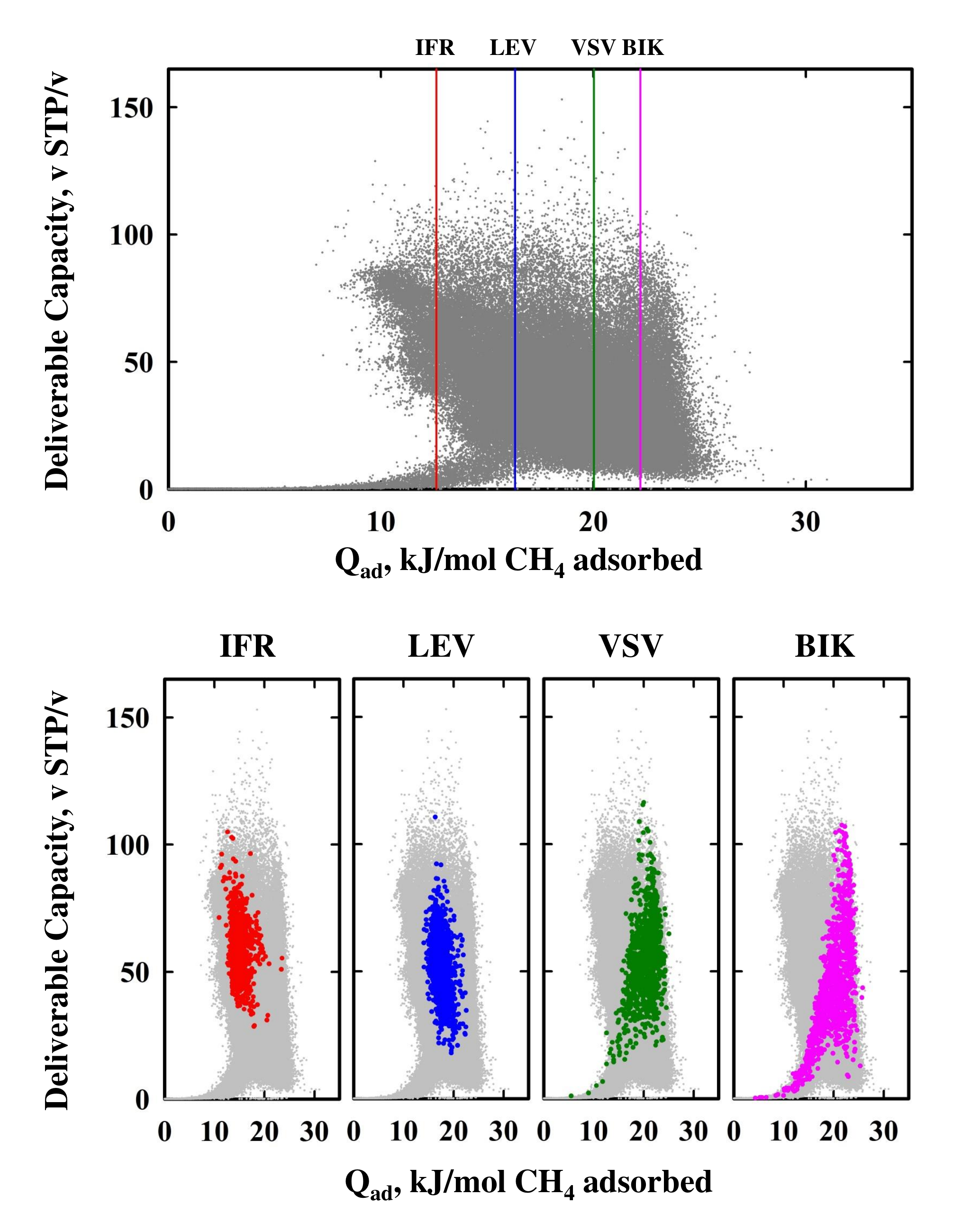}
\caption{}
\label{QadGroups}
\end{figure}

\newpage
\begin{figure}[h!]
\centering
\includegraphics[scale=0.4]{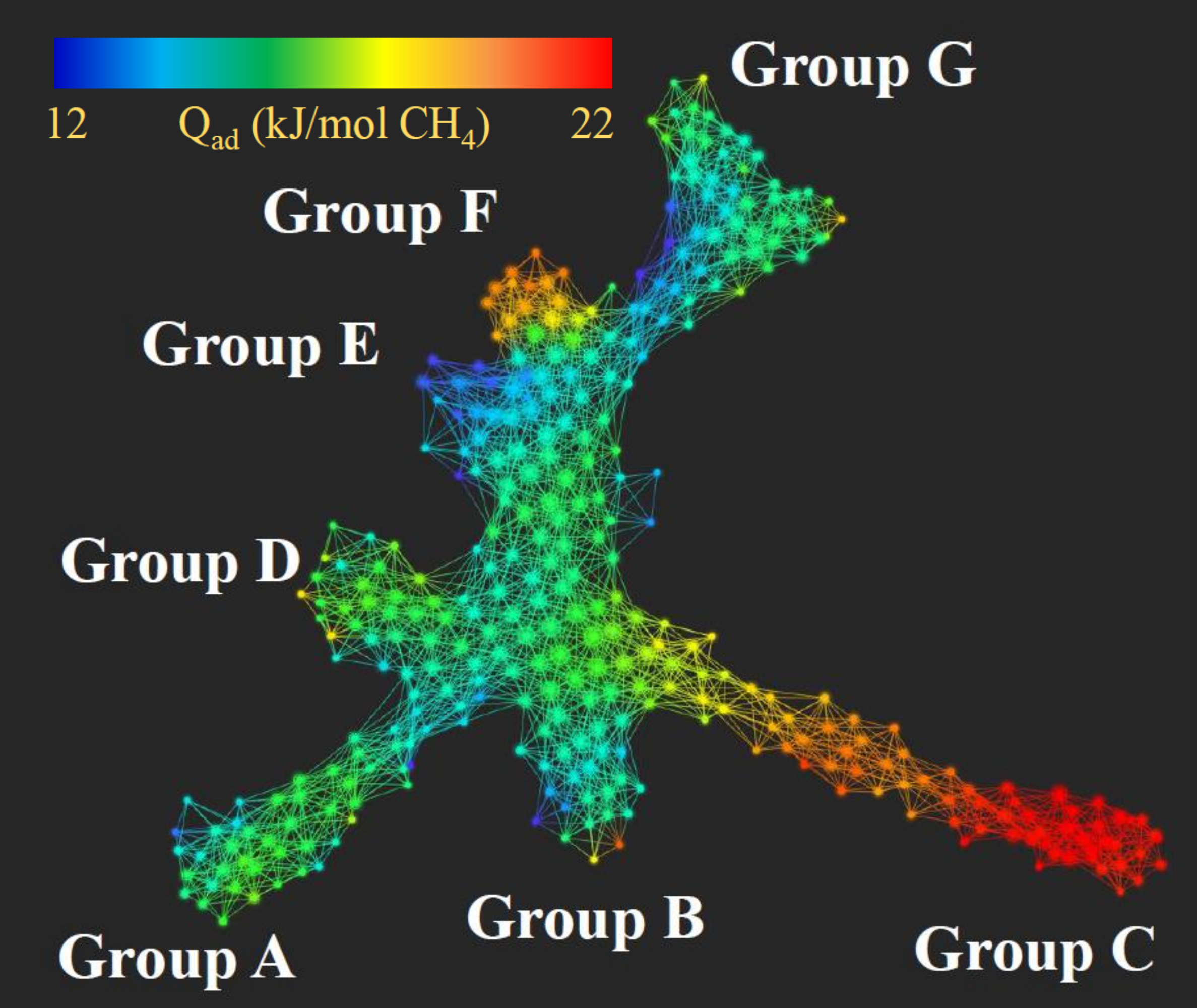}
\caption{ }
\label{mapper}

\end{figure}
\newpage
\clearpage

\clearpage
\setcounter{page}{1}
\begin{center}

{\Large Supplementary Information}\\
~\\
{\large Pore-geometry recognition: on the importance of quantifying similarity in nanoporous materials} \\
Yongjin~Lee, Senja~D.~Barthel, Pawe\l{}~D\l{}otko, S.~Mohamad~Moosavi, Kathryn Hess, and Berend Smit
\\
~\\
correspondence to: berend.smit@epfl.ch
\end{center}
~\\
~\\
{\bf This PDF file includes:}
\begin{itemize}
\item[] Text sections
\begin{itemize}
\item Description of the methods and computational details
\item Additional similarity studies
\item Additional screening studies (global structural properties)
\item Description of the mathematical basis of persistent homology
\item Detailed acknowledgment
\item Additional tables and figures 
\end{itemize}
\item[] Figs. SI-1 to SI-7
\item[] Tables SI-1 to SI-3
\item[] Additional Reference List
\end{itemize}
{\bf Other Supplementary Materials for this manuscript includes the following:}
\begin{itemize}
\item[] Movies S1 to S2
\end{itemize}
\setcounter{figure}{0} 
\renewcommand{\thefigure}{SI-\arabic{figure}} 
\renewcommand{\thetable}{SI-\arabic{table}}

\clearpage
\section{Methods and computational details}\label{methcomp}
\subsection{Methods}\label{methods}
In this section we briefly describe how we assign a descriptor to a porous material using persistent homology.
\vskip 10pt
In order to assign the persistent homological descriptor to a material, we perform the following steps. We start by preparing a supercell of the material by expanding each unit cell to approximately the size of the largest unit cell of all considered materials, in order to compare materials that have  unit cells of very different sizes. The pore system accessible to a gas molecule of interest is determined using the software package Zeo++.~({\it 1}) The surface of this pore system is sampled with a fixed number of points per unit surface area. From these sampled points, filtered Vietoris-Rips complexes are constructed and their 0-, 1-, and 2-dimensional persistence barcodes computed using the software package Perseus.~({\it 2})  We measure the distance between two barcodes by a combination of the $L^{2}$-landscape distances of the barcodes from the dimensions 0,1, and 2, using the Persistence Landscape Toolbox.~({\it 3})

\subsection{Computational details}\label{comp}
 \setcounter{figure}{0} 
The program Zeo++ ({\it 1})  detects the accessible void space inside a porous material using a periodic Voronoi network, modeling the framework atoms as hard spheres with radii taken from the Cambridge Structural Database.~({\it 4,5})  The space accessible to a gas depends on the gas molecule size and is determined in terms of a probe gas molecule, where the size of the probe has to be chosen according to the specific application. We treat a probe gas molecule as a sphere with radius 1.625, 1.5, 1.83, or 1.98~\r{A} for methane, carbon dioxide, krypton, or xenon, respectively. These values are chosen smaller than usual to mimic  by geometric constraints the accessibility of pore space as determined by energy barriers. Zeo++ encodes the pore structure as a large set of points situated on the pore surface which is defined as the boundary of the space where a probe can be placed. For example, a cylinder-shaped pore whose radius equals the probe radius will be represented by points along the central line of the pore. To analyze this point cloud with persistent homology tools, it is necessary to decrease the number of points by performing a secondary sampling, since the raw output is too large to be handled; the raw output is hundreds of thousands of points for each supercell. On the one hand, it is important to have a fine enough resolution to capture details of the pore structures using only finitely many points and to ensure that the barcode assignment is stable with respect to the choice of the point cloud. On the other hand, high resolutions increase computational costs for the persistence computation. We use a combination of random sampling and grid sampling. The grid sampling guarantees that different samplings of a structure give comparable barcodes, in particular by ensuring that points on narrow parts of the pore system  are sampled while still maintaining its connectivity. On the other hand, the random sampling prevents picking up the grid structure in the barcodes. For the random sampling we choose one point per 2~\r{A}$^{2}$ surface area while respecting a minimal distance $r_{\rm min}$ between two sampled points where we decrease $r_{\rm min}$ in steps of 0.1~\r{A} starting with $r_{\rm min}=1.3$~\r{A} until the given number of points has been selected.  The grid size is 0.5~\r{A} and for each cube of the grid the point of the original point cloud is chosen that is closest to the midpoint of the cube. A point of the grid sampling is added to the random sampling whenever its distance to the randomly sampled points is greater than the final value of $r_{\rm min}$. 

\begin{figure}[t]
\begin{center}
\vskip -2cm
\includegraphics[scale=0.5]{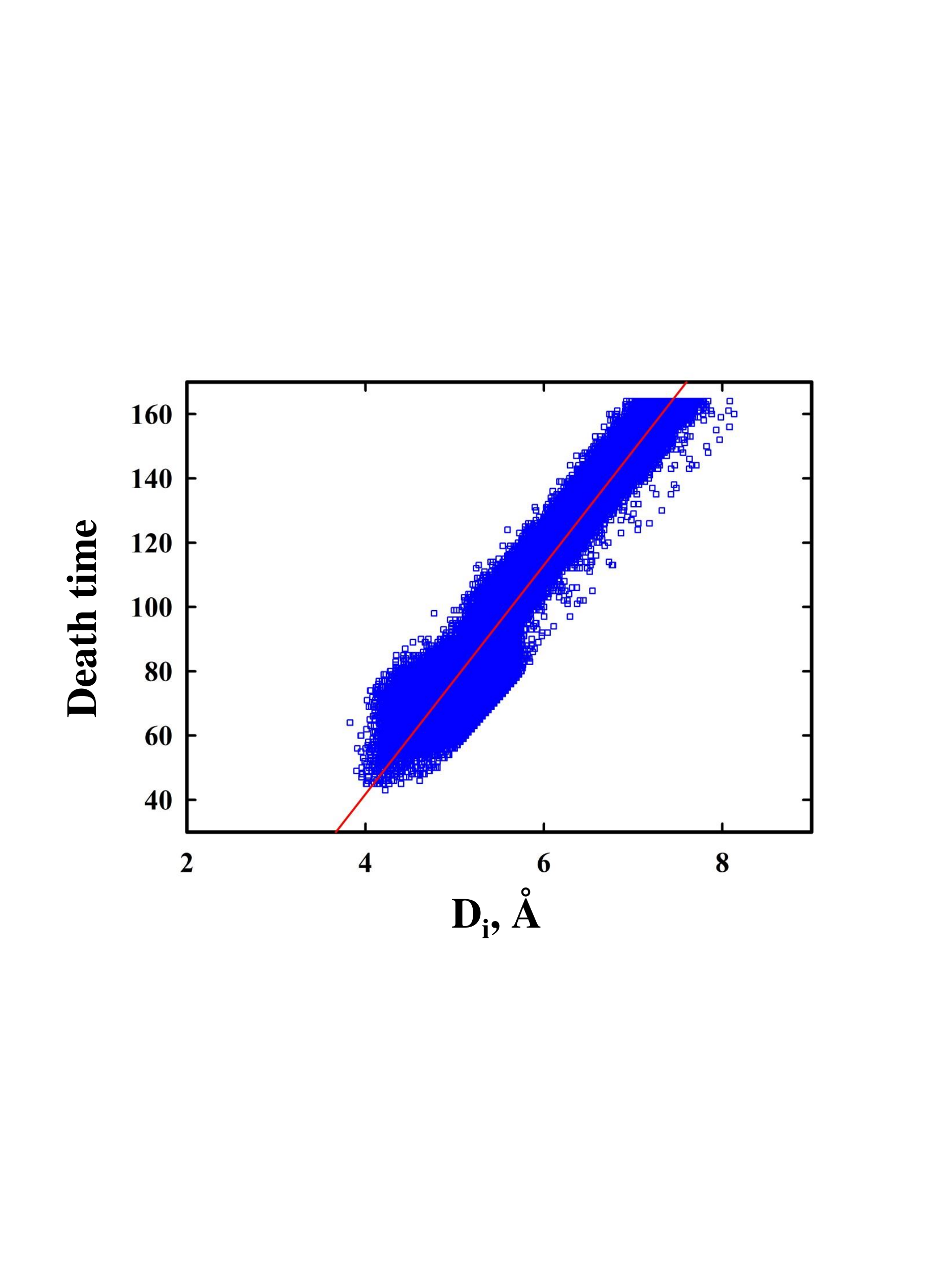} 
\vskip -3cm
\caption{\label{PH2_CH4} Correlation of the death time of 2-dimensional homology classes and the diameters of the largest included sphere $D_i$ when using methane CH$_{4}$ as a probe molecule. The red line indicates the least squares regression line; Death time = 35.6$\times$$D_i$--100.8.}
\end{center}
\end{figure}

The second step towards the persistent homological descriptor consists of calculating the persistence barcodes for a filtration of Vietoris-Rips complexes, obtained from the sets of points computed in the first step using the software package Perseus.~({\it 2}) We restrict ourselves to constructing \mbox{3-dimensional} Vietoris-Rips complexes, where the filtration parameter~$\epsilon$ (corresponding to the radius of the balls grown around each point) increases in 164 steps of 0.025~\r{A} increments, starting from the initial value of 0. The resulting 4.1~\r{A} maximal filtration parameter is due to the fact that the memory cost needed by Perseus grows extremely fast as the radius increases in our calculations. While the relatively small maximal filtration parameter does not allow us to build a complete complex, it prevents geodesically distant points of the surface that are close in Euclidean metric to be connected unless the pore structure is very densely packed in the material. This is important since our construction does not distinguish homology classes that are formed in the solid part of the material from those formed in the pore regions. Technically, this makes the descriptor an overall descriptor of the geometry of the embedding of the pore-surface in the ambient space and not strictly describing the pore surface with respect to the pore space only. Fortunately, the technique does not tend to misidentify structures since the material part is typically much larger than the pore part. However, our maximal filtration parameter is not sufficiently large for all homology classes to die -- these correspond to essential intervals in the barcodes -- especially for zeolites having large pores. Therefore, to take account of these homology classes in computing distances between two barcodes, a maximal value for the death time has to be assigned, which is especially important in dimension 2 because of the small cardinality of barcodes. For 2-dimensional barcodes, we assign a death times to essential intervals based on the relation between the diameter $D_{i}$ of the largest included sphere and the death time for small and medium pores which is linearly fitted. An example for zeolites with methane is shown in Figure~\ref{PH2_CH4}. The 1-dimensional barcodes contain sufficiently many intervals to distinguish different structures and we discard essential intervals.

To quantify the similarity between two materials in the barcode space, we combine as follows the $L^2$-distances between the persistence landcapes (see section~\ref{theo}) corresponding to their barcodes in the different dimensions. After testing landscape distances of different orders (i.e. $L^\infty$, $L^0$, $L^1$, $L^2$, and so on),  $L^2$-distances were chosen because they gave the smallest errors in predicting global structural properties and performance properties for a test set of materials. Let $\Lambda_{d=1}$ (resp. $\Lambda_{d=2}$) be the $L^2$-landscape distance between the 1-dimensional (respectively, 2-dimensional) persistence barcodes,  and let  $L_{0}=\left|\frac{n_{1}}{V_{1}}-\frac{n_{2}}{V_{2}}\right|$ where $n_{i}$ is the number of points sampled on the pore surface of the $i^{\text{th}}$ material, and $V_{i}$ is the volume of the supercell. The distance between two materials in the barcode space is then
$$D_{TS}:= \sqrt{\alpha_{0}L_{0}^{2}+\alpha_{1}\Lambda_{d=1}^{2}+\alpha_{2}\Lambda_{d=2}^{2}},$$ 
with coefficients $\alpha_{0}= 0.1, \; \alpha_{1}=0.45$, and $\alpha_{2}=0.45$, the values of which were chosen to minimize the error in predicting global structural properties and performance properties for a test set of materials. In dimension 0 the essential intervals are effectively discarded and instead the 0-dimensional barcode the number of sampled points per unit volume is used. This is a simplification that corresponds to discarding the essential intervals in all cases where different connected components of the pore system stay separated during the entire filtration; the 0-dimensional barcodes of connected components are determined by the sampling procedure by construction.

The distance $D_{CS}$ between two materials in the conventional descriptor space is estimated with a normalized euclidean distance of five conventional structural properties with an equal weigh for each: $D_i$ (the diameter of largest included sphere), $D_f$ (the diameter of largest free sphere), $\rho$ (density of a framework), $ASA$ (accessible surface area), and $AV$ (accessible volume). 
\clearpage
\section{Similarity in different classes of nanoporous materials}\label{similarity}

To illustrate the application of our method to finding similar pore geometries across different classes of nanoporous materials, we consider the following questions. 
\begin{enumerate}
\item Are there zeolites that have the same pore geometry as a given MOF? 
\item Are there hypothetical MOFs that are similar to MOFs that have already been synthesized? 
\item Are there materials among the ca.~5000  MOF structures that are deposited in the Cambridge Structural Database~({\it 6}) that have similar pore structures but different chemical compositions? 
\end{enumerate}
The common theme behind these questions is to illustrate how the methodology developed here allows researchers to identify materials that have similar pore geometries.
 
\subsection{MOFs and zeolites}\label{mofzeol} 
\begin{figure}[h!]
\begin{center}
\vskip -4cm
\includegraphics[scale=0.5]{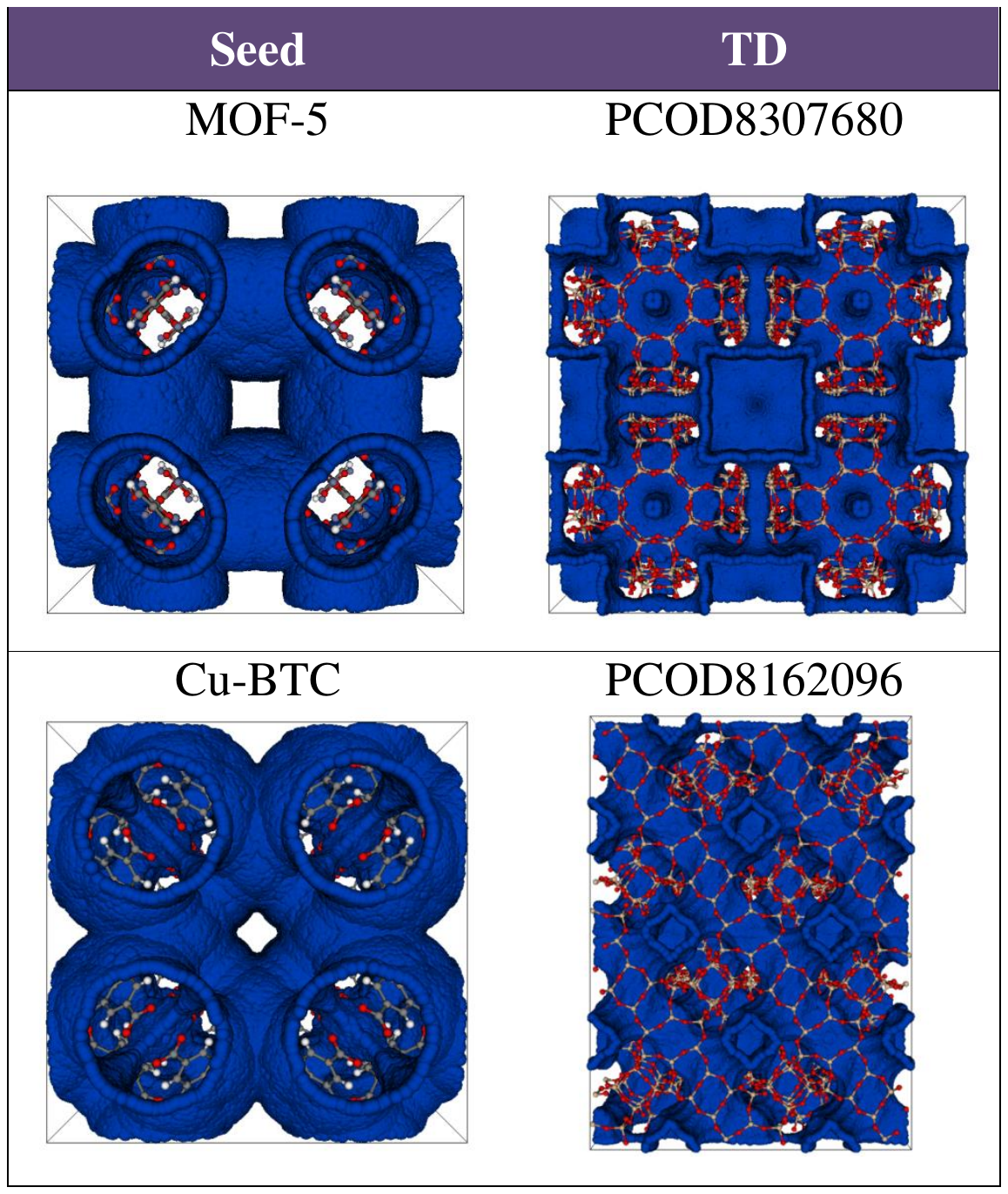} 
\vskip -4cm
\caption{\label{MOFzeo}The zeolites most similar to MOF-5 and Cu-BTC with respect to PerH.}
\end{center}
\end{figure}

In Figure~\ref{MOFzeo}, we identify the structures in the IZA+ hypothetical zeolite database~({\it 7,8}) that are most topologically similar to some of the best known MOFs (e.g., MOF-5 and Cu-BTC). The figure shows that we can indeed find hypothetical zeolite structures that look very similar to these MOFs.

\subsection{Hypothetical and experimental MOFs} \label{hmof}
\begin{figure}[h!]
\begin{center}
\includegraphics[scale=0.5]{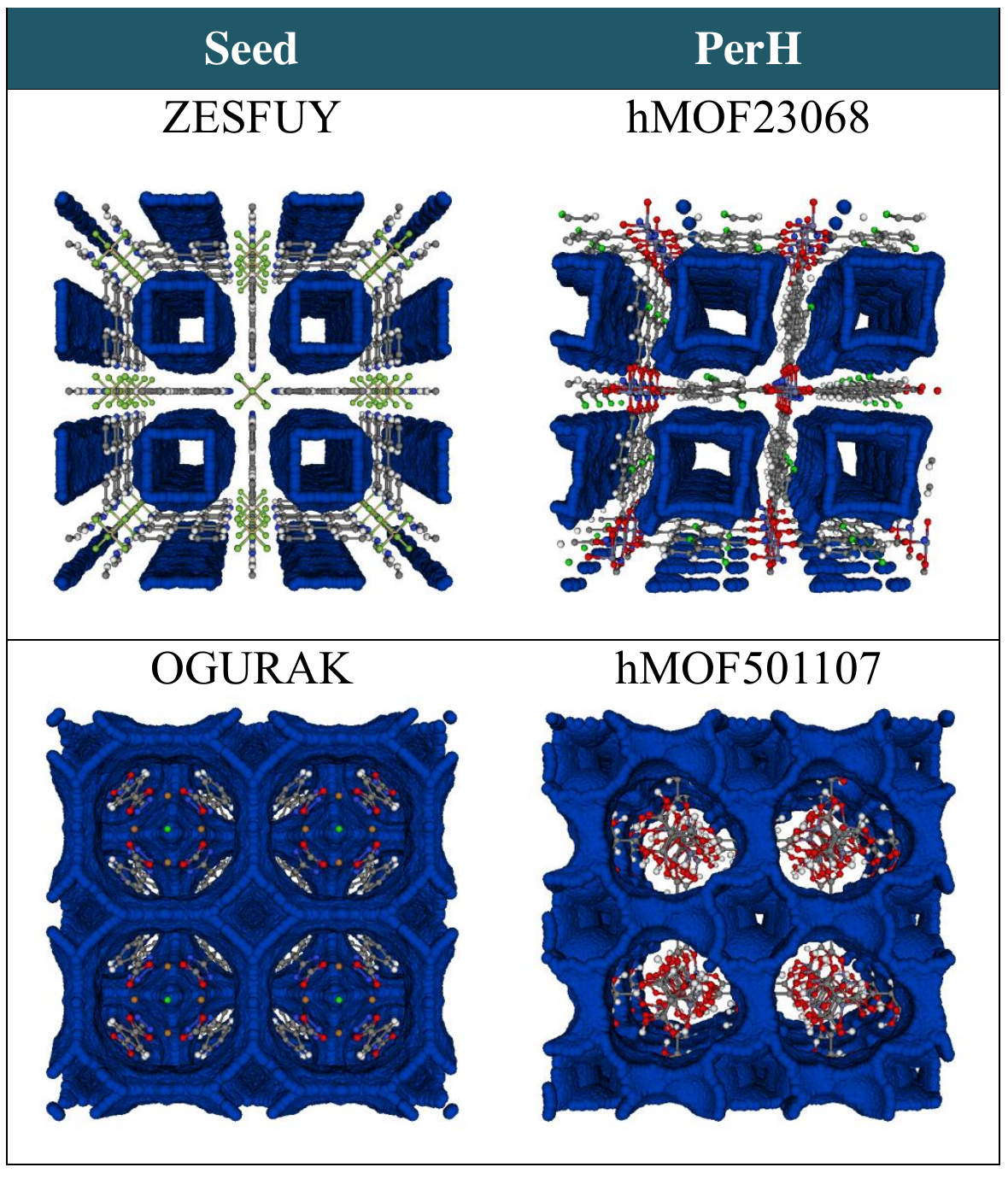} 
\caption{\label{hMOFexMOF}Finding hypothetical MOFs that best resemble the experimentally known structures ZESFUY and OGURAK.}
\end{center}
\end{figure}

For hypothetical MOFs we have a database of over 140,000 materials.~({\it 9}) An interesting question is whether  pore geometries similar to those occurring in hypothetical MOFs have already been synthesized. This question is difficult to answer  with traditional methods, since the materials might differ in their chemical composition. We have compared the similarity of structures from the database of hypothetical MOFs (hMOFs) with the experimental structures in the CoRE-MOF database. Figure~\ref{hMOFexMOF} shows two examples of similar structures. The color coding of the structures shows that the chemical composition of the two structures is very different.

\clearpage
\section{Global structural properties}\label{screening}

In the main text we explained that the different dimensions of the persistent homology of the structures that we consider admit geometric interpretation. It is therefore interesting to see whether we can detect this geometric content  when we use our method to test the capability of PerH to screen zeolites for the following conventional structural properties: $D_i$, $D_f$, $\rho$, $ASA$, $AV$, the Henry coefficient ($K_H$), and the heat of adsorption ($Q_{ad}$). We use methane as a probe molecule.

Starting with a highly diverse training set of 600 structures chosen by the min-max algorithm ({\it 10}) , we perform high-throughput screening for the entire set of zeolites, using five different PerH's: $PD_{0}$ (=$L_{0}$ as defined in~\ref{methods}), $PD_1$, and $PD_2$, which use only 0-, 1-, or 2-dimensional persistent homology information respectively, as well as $PD_{12}$ and $PD_{012}$ which combine information from 1- and 2- or 0-, 1-, and 2-dimensional persistent homology. For $PD_{12}$ equal weights were used and  for $PD_{012}$ the same weights described previously. For each screening, we compare the conventional properties of each zeolite with those of the most similar one in the training set. Table~\ref{BarcodeErrors} summarizes the mean absolute percentage errors (MAPE) for each property which is calculated as 
$$\text{MAPE}= \frac{1}{n}\sum_{i=1}^{n} \left | \frac{PP_{i,PerH}-PP_{i}}{PP_{i}}\right |,$$ 
where $n$ is the number of zeolites in the promising set (as defined in~\ref{methods}), and $PP_{i}$ (respectively, $PP_{i,PerH}$) is the performance property of the $i^{\text{th}}$ zeolite (respectively,  of the zeolite in the training set most similar to the $i^{\text{th}}$ zeolite). 

We observe that TDA-based descriptors are also capable of screening for structural properties. Moreover, the different dimensions of the persistent homology detect different structural properties. Using the standard deviation of the prediction of the screening as a measure of the quality of the description, Table~\ref{BarcodeErrors} shows that the best prediction for the surface area (ASA) is the 0-dimensional descriptor ($PD_0$), while the maximum included sphere ($D_i$) is best predicted with a \mbox{2-dimensional} descriptor  ($PD_2$). In section \ref{theo} we explain that this corresponds to the geometric interpretation of the persistent homology in the different dimensions. In addition Table~\ref{BarcodeErrors} shows that averaging over the three dimensions provides a good description of all properties. 

\begin{table}[h!]
\begin{center}
\begin{tabu}{l|lllll}
Property                          & $PD_{0}$ & $PD_1$ & $PD_2$ & $PD_{1,2}$ & $PD_{0,1,2}$ \\ \hline
$K_{H}^{\ast}$ & 0.080     & 0.087 & 0.074 & 0.085       & 0.086         \\
$\rho$                              & 0.082     & 0.107 & 0.121 & 0.096       & 0.073         \\
$Q_{ad}$                      & 0.369     & 0.392 & 0.412 & 0.379       & 0.386         \\
$ASA$                           & 0.078     & 0.476 & 0.621 & 0.459       & 0.091         \\
$D_{i} $                         & 0.318     & 0.367 & 0.155 & 0.234       & 0.172         \\
$D_f$                              & 0.346     & 0.263 & 0.344 & 0.293       & 0.158         \\
$AV^{\ast}$       & 0.217     & 0.328 & 0.319 & 0.312       & 0.194              
\end{tabu}\vskip 3pt
$^{\ast}$Mean absolute percentage errors of $K_{H}$ and $AV$ are obtained with $\log K_{H}$ and $\log AV$.
\vskip 5pt
\end{center}
\caption{TDA analysis of the conventional descriptors $K_H$ (Henry coefficient), $Q_{ad}$ (heat of adsorption), $\rho$ (density), $D_{i}$ (maximum included sphere), $D_{f}$ (maximum free sphere), $ASA$ (accessible surface area), and $AV$ (accessible volume). The data show the mean absolute percentage error, expressing how well these descriptors can be predicted on the basis of a training set using only the 0-D, 1-D or 2-D barcode as fingerprint, ($PD_0$, $PD_1$, and $PD_2$, respectively), the combined 1-D and 2-D barcodes ($PD_{1,2}$), and the combination of barcodes from all 3 dimensions ($PD_{0,1,2}$). The mean absolute percentage error is calculated as $\frac{1}{n}\sum_{i=1}^{n}{\left| \frac{P_{i,PD}-P_{i}}{P_{i}} \right|}$ where $n$ is the number of zeolites, and $P_i$ or $P_{i}$, $P_{i,PerH}$ is a property of a zeolite or that of the most similar zeolite selected by $PD$, respectively.}
\label{BarcodeErrors}
\end{table}

\clearpage
\section{Theoretical background}\label{theo}

Topological data analysis (TDA) is a mathematical technique for assigning various topological invariants to data. The guiding philosophy of TDA is that the `shape' of the data, encoded by a mathematical `signature', should reveal important relations among the data points. One of the best-known TDA techniques is persistent homology$^{11,12}$, which we describe very briefly here. Let us denote that persistent homology have already been used in material analysis to detect various phase states of glass.$^{13}$ 

Each material is encoded as a point cloud obtained by sampling points on the pore surface, giving rise to a description of the material in terms of the coordinates of the sampled points in 3-space. From the points, we construct a filtration of Vietoris-Rips complexes, which is a sequence of nested triangulated objects. 

For a fixed non-negative real number $r$, the Vietoris-Rips complex of a point cloud is constructed  as follows from the collection of balls of  radius $r$ centered at the points of the point cloud. Starting with the elements of the point cloud, add a line segment between a pair of points when the balls centered at the two points overlap. Similarly, a solid triangle is added when each pair of balls centered at its corners intersect and a solid tetrahedron when four balls all intersect pairwise. This procedure can be extended to all higher dimensions, but we stopped at solid tetrahedra both for computational reasons and because our point cloud represented a real three-dimensional structure. Since the Vietoris-Rips complex for a small radius is contained in the complex for a bigger radius, we obtain a filtration of complexes (Figure~\ref{Complex} top).

The shape of a complex is partly captured by its homology groups $H_{n}$, where $n$ is a non-negative integer. The nonzero elements of  $H_{n}$ are the homology classes in dimension $n$, which correspond to the $n$-dimensional `holes' in the complex. More precisely, the 0-dimensional homology classes correspond to the connected components, while a 1-dimensional homology class is represented by a closed curve that does not bound a surface and a 2-dimensional homology class  by a bounded cavity.

\begin{figure}[h]
	\centering
	\def\svgwidth{400pt}
 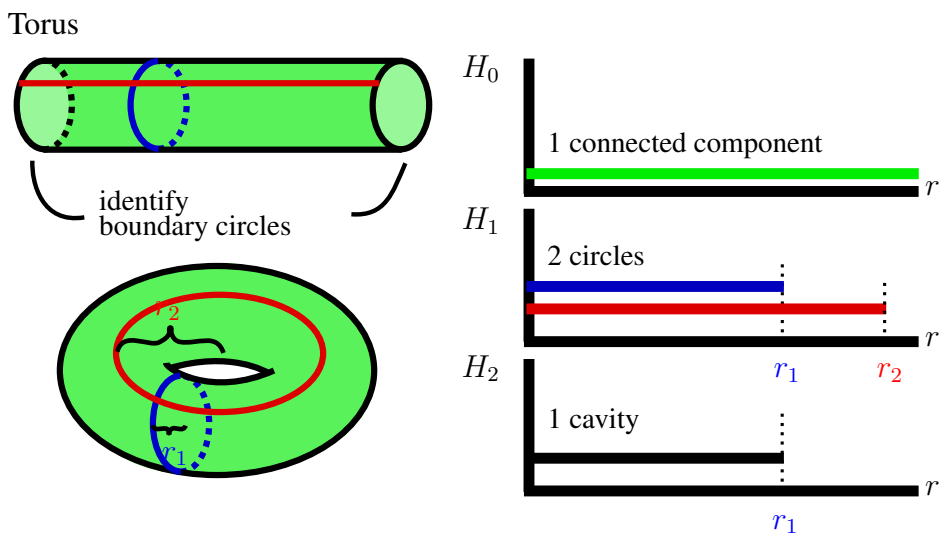
\caption{The persistence barcodes of a torus as obtained from a channel by implying periodic boundary conditions.}
\label{homologies}
\end{figure}

For example, a hollow tube has one 0-dimensional homology class since it is connected, one 1-dimensional homology class corresponding to  the circle running around the axis of the tube, which does not bound a disk, and no 2-dimensional homology class, as the tube does not bound a 3-dimensional cavity. In contrast, if the ends of the tube are glued together, for example by applying periodic boundary conditions, a torus is formed (see Figure~\ref{homologies}). Being connected, it still has one 0-dimensional homology class, but two independent 1-dimensional homology classes, which are represented by the circle around the axis (blue) and the newly formed circle that runs parallel to the axis (red). The torus is hollow and thus bounds a cavity that represents a  2-dimensional homology class. If a solid torus is considered, the circle around the axis bounds a disk and therefore does not contribute to the 1-dimensional homology, and there is no nonzero 2-dimensional homology class.

The homology classes of a point cloud (such  as that obtained by sampling a pore surface) are not very informative, since each point forms its own connected component, while $H_{n}=0$ for all $n>0$. In contrast, the homology groups of its Vietoris-Rips complexes strongly depend on the position of the points in space. This information is stored in persistence barcodes that track the non-trivial homology classes through the radius-dependent filtration. A persistence barcode is a set of intervals where each nontrivial homology class is represented by a bar. The starting point of an interval denotes the smallest radius for which the homology class represented by the interval (e.g., a circle around a hole in dimension~1) appears in homology of the associated Vietoris-Rips complex, while the endpoint is given by the radius where the homology class disappears (e.g., the smallest radius for which the balls close the hole) (Figure~\ref{Complex} bottom, Figure~\ref{homologies}). Classes that have a short lifetime can be considered as noise, while classes that persist through long intervals reveal actual structural features of the point cloud. 

\begin{figure}[h]
	\centering 
	\def\svgwidth{300pt}
 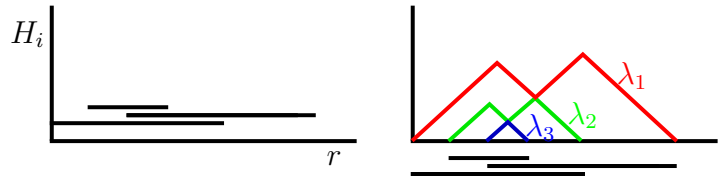
\caption{A persistence barcode and its corresponding persistence landscape.}
\label{landscape}
\end{figure}

\begin{figure}[h!]
	\centering
	\def\svgwidth{460pt}
 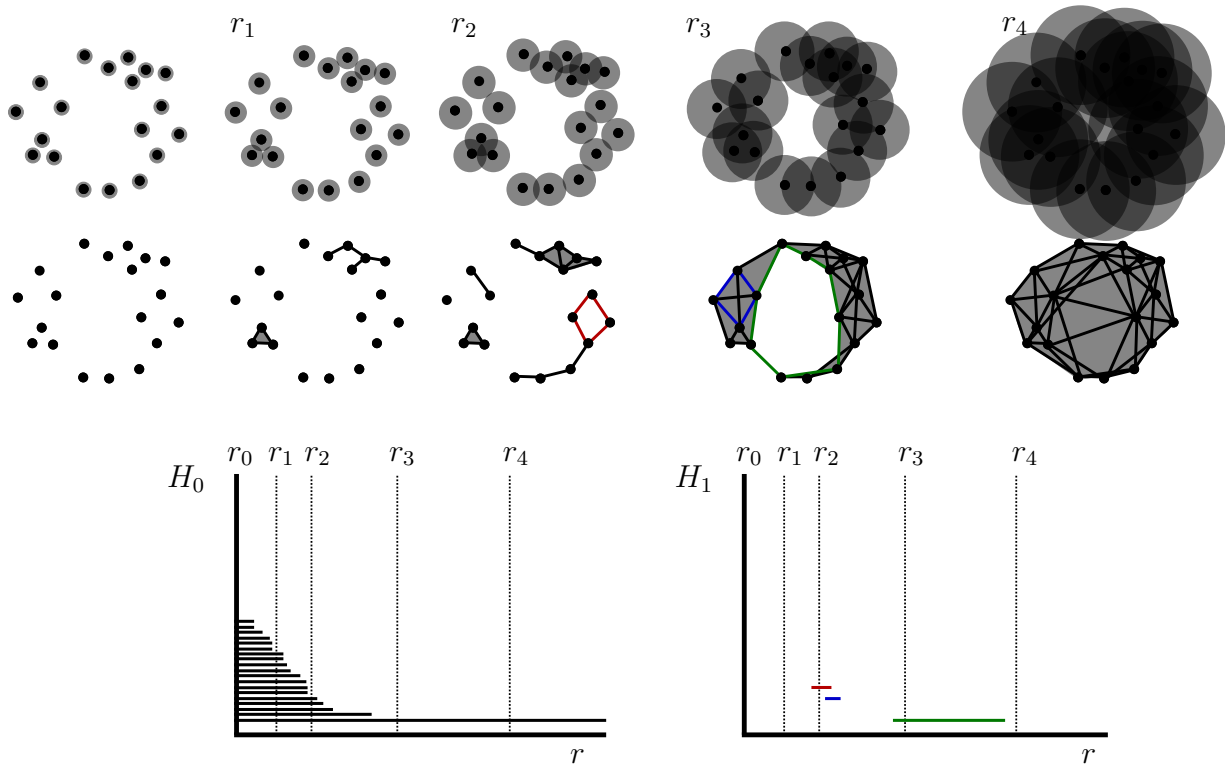
\caption{Construction of the Vietoris-Rips complex from a point cloud in 2D for increasing radii, together with the  0- and 1-dimensional persistence barcodes of the resulting filtration. $H_{0}$ counts the connected components of the complex for a given radius, and $H_{1}$ tracks circles that do not bound disks. The construction in 3D works analogously, using balls instead of disks.}
\label{Complex}
\end{figure}

To compare two materials in terms of their persistence barcodes, we use a combination of the $L^2$-distances between the persistence landcapes corresponding to the persistence barcodes of same dimensions. Informally, a persistence landscape is a family of functions 
$$\lambda =\big\{\lambda_{k} : \mathbb{R} \rightarrow \mathbb{R} \cup \{\infty \}\mid k\in \mathbb N\big\}$$ 
obtained from a barcode by ``stacking together isosceles triangles whose bases are the intervals of the barcode,'' where the $k^{\text{th}}$ function describes the contour of the $k^{\text{th}}$ maximum (Figure~\ref{landscape}); see {\bf Ref} 14 for a rigorous definition. The $L^2$-landscape distance between two persistence barcodes $B$ and $B'$  with corresponding persistence landscapes $\lambda$ and $\lambda '$  is given by 
$$\Lambda(B,B') = \| \lambda - \lambda'\|_{2}=\sum_{k=1}^{\infty}\left( \int | \lambda_{k}(t)-\lambda'_{k}(t)|^{2}dt\right)^\frac{1}{2}$$

\newpage

\newpage

\section*{Detailed acknowledgment}
During the early stage of the research YL and BS were supported by the Center for Gas Separations Relevant to Clean Energy Technologies, an Energy Frontier Research Center funded by the DOE, Office of Science, Office of Basic Energy Sciences under award DE-SC0001015.  
 
YL (during the later stages of the research) and SB were supported by the National Center of Competence in Research (NCCR) ``Materials' Revolution: Computational Design and Discovery of Novel Materials (MARVEL)'' of the Swiss National Science Foundation (SNSF).

MM was supported by the Deutsche Forschungsgemeinschaft (DFG, priority program\\SPP 1570).
 
BS was supported by the European Research Council (ERC) under the European Union's Horizon 2020 research and innovation program (grant agreement No 666983).  

PD was supported by the Advanced Grant of the European Research Council 
GUDHI, (Geometric Understanding in Higher Dimensions) (grant agreement No 339025). 

\newpage

\section*{Additional figures and tables SI}


\begin{figure}[h!]
\centering
\includegraphics[width=\textwidth]{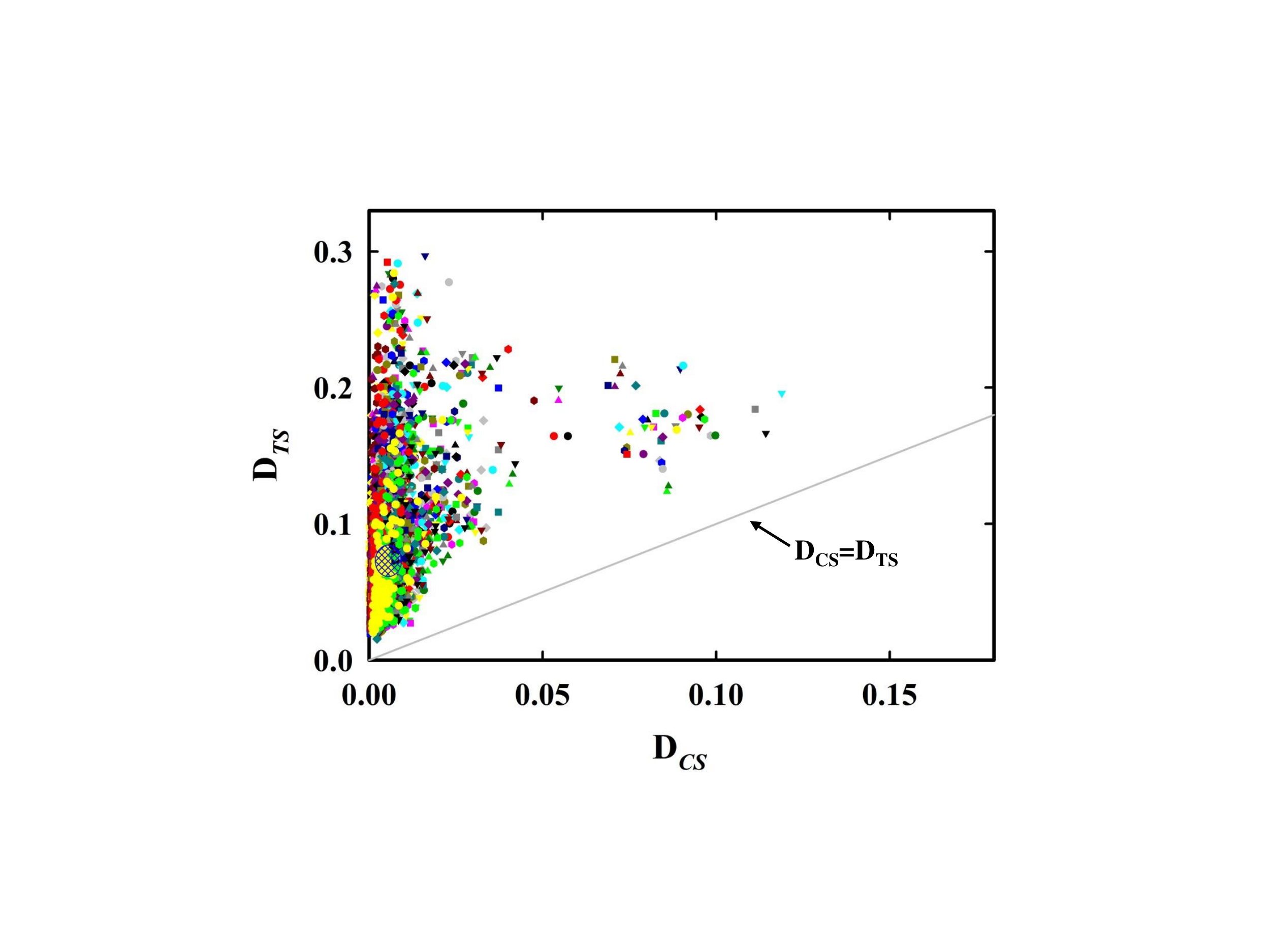}
\caption{The effect of weighting the conventional properties differently on ConD. The x and y axes give the average of the distances ($D_{CS}$ or $D_{TS}$) from four most similar materials to the corresponding reference zeolite structure (seed structure for searching similar ones) measured in conventional or barcode space respectively. 50 different combinations of weight factors were chosen randomly, and the results for each combination are distinguished using different colors. A cross-hatched ellipse shows the area that contains the centers of the point clouds which are obtained from different weighting choices.}
\label{Weight_Effects}
\end{figure}

\newpage
\begin{table}
\vskip -4cm
\begin{centering}
\includegraphics[scale=0.85]{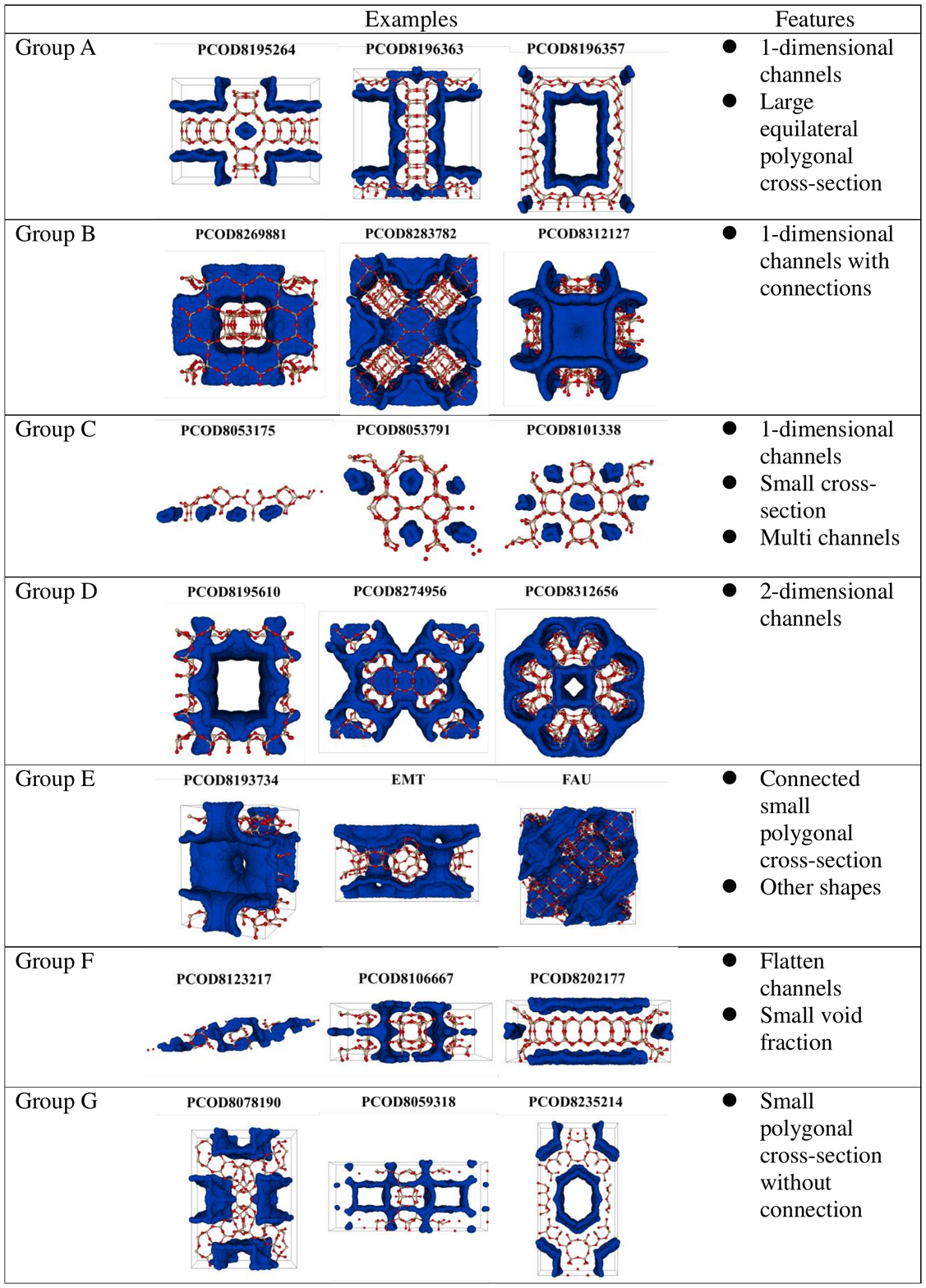}
\vskip -2cm
\end{centering}
\caption{Examples from the seven topologically different classes of top-performing materials  for methane storage (see also Figure~4).  }
\label{Groups} 
\end{table}
\newpage

\begin{table}
\begin{tabu}{l|[1.5pt] cclllll}
Descriptor				&		&	 name		&	$D_{i}$	& $D_{f}$ 	& 	$\rho$ 	& 	$ASA$		&	 $AV$	\\  \tabucline[1.5pt]
					&	& Seed	& 	SSF		&	7.59		&	6.15	&	1.64	&	1191.97	&	0.122	\\
						& 1st   	&	PCOD8328603	&	8.09 		& 	6.34	&	1.77	&	1167.86	&	0.120	\\
\textcolor{blue}{PerH}	& 2nd	&	PCOD8267032	&	7.54		&	6.22	&	1.70	&	1171.01	&	0.115	\\
						& 3rd	&	PCOD8267258	&	7.72		&	6.21	&	1.63	&	1205.27	&	0.115	\\
						& 4th	&	PCOD8325065	&	7.63		&	6.29	&	1.59	&	1239.91	&	0.133	\\ \hline
						& Seed	& 	SSF		&	7.59		&	6.15	&	1.64	&	1191.97	&	0.122	\\
						& 1st   	&	PCOD8242590	&	7.87 		& 	6.16	&	1.62	&	1210.05	&	0.119	\\
\textcolor{red}{ConD}	& 2nd	&	PCOD8239380	&	7.60		&	6.29	&	1.63	&	1156.76	&	0.120	\\
						& 3rd	&	PCOD8267258	&	7.72		&	6.21	&	1.63	&	1205.27	&	0.115	\\
						& 4th	&	PCOD8070132	&	7.69		&	6.49	&	1.62	&	1187.66	&	0.126	\\ \tabucline[1pt]
					&	& Seed	& 	IWV		&	8.48		&	6.97	&	1.50	&	1502.63	&	0.181	\\
						& 1st   	&	PCOD8285528	&	8.46		& 	6.86	&	1.54	&	1476.78	&	0.176	\\
\textcolor{blue}{PerH}	& 2nd	&	PCOD8329417	&	7.88		&	6.65	&	1.49	&	1507.05	&	0.189	\\
						& 3rd	&	PCOD8284133	&	9.17		&	6.92	&	1.63	&	1491.56	&	0.194	\\
						& 4th	&	PCOD8283909	&	9.08		&	6.60	&	1.49	&	1534.93	&	0.187	\\ \hline
						& Seed	& 	IWV		&	8.48		&	6.97	&	1.50	&	1502.63	&	0.181	\\
						& 1st   	&	PCOD8302674	&	8.78 		& 	7.00	&	1.50	&	1499.77	&	0.183	\\
\textcolor{red}{ConD}	& 2nd	&	PCOD8310713	&	8.39		&	7.05	&	1.49	&	1533.70	&	0.178	\\
						& 3rd	&	PCOD8079814	&	8.33		&	7.07	&	1.48	&	1523.32	&	0.175	\\
						& 4th	&	PCOD8059487	&	7.99		&	7.01	&	1.48	&	1506.13	&	0.180						
\end{tabu}
\caption{The global structural properties of the four zeolites most similar to SSF and IWV selected by either conventional descriptors (ConD) or using persistance homology (PerH):  $D_{i}$ (maximum included sphere),  $D_{f}$ (maximum free sphere),  $\rho$ (density), $ASA$ (accessible surface area), and $AV$ (accessible volume).  }
\label{StructProp}
\end{table}

\clearpage
\newpage
\section*{References}
\baselineskip14pt
\begin{enumerate}

\item Martin R. L., Smit B. \& Haranczyk M., Addressing challenges of identifying geometrically diverse sets of crystalline porous materials. {\it J. Chem. Inf. Model.\/} {\bf 52}, 3078--318 (2012).

\item Mischaikow K. \& Nanda V., Morse theory of filtrations and efficient computation of persistent homology. {\it Discrete Comput. Geom.\/} {\bf 50}, 330--353 (2013).

\item Bubenik P. \& D\l{}otko P., A persistence landscapes toolbox for topological statistics. {\it J. Symbolic Comput. \/} {\it in press \/} (2016).

\item Bondi A., Van der Waals volumes and radii. {\it J. Phys. Chem.\/} {\bf 68}, 441--451 (1964).

\item Rowland R. S. \& Taylor R., Intermolecular Nonbonded Contact Distances in Organic Crystal Structures: Comparison
with Distances Expected from van der Waals Radii. {\it J. Phys. Chem.\/} {\bf 100}, 7384--7391 (1996).

\item Chung Y. G., Camp J., Haranczyk M., Sikora B., Sholl D. S. \& Snurr R. Q., Compilation and Screening of a Database of Computation-Ready Experimental (CoRE) Metal-Organic Frameworks. {\it Chem. Mater.\/} {\bf 26}, 6185--6192 (2014).

\item Pophale R., Cheeseman P. A. \& Deem M. W., A database of new zeolite-like materials. {\it Phys. Chem. Chem. Phys.\/} {\bf 13}, 12407--12412 (2011).

\item Deem M. W., Pophale R., Cheeseman P. A. \& Earl D. J., Computational discovery of new zeolite-like materials. {\it J. Phys. Chem. C.\/} {\bf 113}, 21353--21360 (2009). 

\item Wilmer C. E., Leaf M., Lee C. Y., Farha O. K., Hauser B. G., Hupp J. T. \& Snurr R. Q., Large-scale screening of hypothetical metal–organic frameworks. {\it  Nat. Chem.\/} {\bf 4}, 83--89 (2012).

\item Kennard R. W. \& Stone L. A., Computer aided design of experiments. {\it Technometrics\/} {\bf 11}, 137-148 (1996). 

\item Edelsbrunner H. \& Harer J. L., {\it Computational Topology: an introduction\/} (American Mathematical society, Providence RI, 2010).

\item Carlsson G., Topology and data. {\it Cull Amer. Math. Soc.\/} {\bf 46}, 255--308 (2009).

\item Hiraoka Y., Nakamura T., Hirata A., Escobar E.G.. Matsue K. \& Nishiura Y., Hierarchical structures of amorphous solids characterized by persistent homology. {\it Proc. Natl. Acad. Sci. USA \/} {\bf 113}, 7035--7040 (2016).

\item Bubenik P., Statistical topological data analysis using persistence landscapes. {\it J. Mach. Learn. Res.\/} {\bf 16}, 77--102 (2015).

\end{enumerate}

\end{document}